\documentclass{article}

\PassOptionsToPackage{numbers,compress}{natbib}

\usepackage[main, final]{neurips_2026}

\usepackage[utf8]{inputenc} % allow utf-8 input
\usepackage[T1]{fontenc}    % use 8-bit T1 fonts
\usepackage{hyperref}       % hyperlinks
\usepackage{url}            % simple URL typesetting
\usepackage{booktabs}       % professional-quality tables
\usepackage{amsfonts}       % blackboard math symbols
\usepackage{nicefrac}       % compact symbols for 1/2, etc.
\usepackage{microtype}      % microtypography
\usepackage{xcolor}         % colors
\usepackage{graphicx}       % figures
\usepackage{xspace}
\usepackage{tipa}           % pronunciation 
\usepackage{booktabs}       % for professional tables
\usepackage{wrapfig}
\usepackage{amsmath}        % Math
\usepackage{amssymb}
\usepackage{mathtools}
\usepackage{amsthm}
\usepackage{bbm}
\usepackage{multirow}
\usepackage[table]{xcolor}
\usepackage{booktabs}
\usepackage{xcolor}
\usepackage{caption}
\usepackage{subcaption}
\usepackage{tcolorbox}
\usepackage{makecell}
\usepackage{arydshln}

\theoremstyle{plain}

\theoremstyle{definition}

\theoremstyle{remark}

\usepackage{xspace}
\usepackage{xcolor}        % for \textcolor
\usepackage{algorithm}
\usepackage{algorithmic}
\usepackage{multirow}
\usepackage{siunitx}
\usepackage{bbm}
\newcommand{\papertablefont}{\footnotesize} % slightly increased font size from 7pt -> 8pt

\newcommand{\ours}{QUTCC\xspace}
\title{\ours \raisebox{-2px}{\includegraphics[scale=0.12]{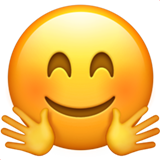}}: Quantile Uncertainty Training and Conformal Calibration for Imaging Inverse Problems}

\author{%
  Cassandra Tong Ye \\
  \texttt{cassye@cs.cornell.edu} \\
  \and
  Shamus Li \\
  \texttt{shamus@cs.cornell.edu} \\
  \and
  Tyler King \\
  \texttt{ttk22@cornell.edu} \\
  \and
  Kristina Monakhova \\
  \texttt{monakhova@cornell.edu} \\
}

\begin{document}

\maketitle

\begin{abstract}
    While deep learning offers tremendous promise for scientific and medical imaging, any failures and hallucinations (predictions that do not coincide with reality) are hard to pinpoint and can have serious downstream consequences. Uncertainty estimation techniques, such as conformal prediction, can help by predicting statistically valid error bars for a model's prediction. However, popular conformal prediction methods were not designed for high-dimensional image-valued problems and do not take into account spatial correlations within an image during conformal calibration, resulting in larger-than-necessary uncertainty intervals. We propose a practical simultaneous quantile regression method that enables non-linear, spatially-adaptive scaling during conformal calibration. Our method, \ours uses a U-Net architecture with a quantile embedding to learn a full conditional quantile distribution during training, and then leverages this non-linear, learned function for \textit{spatially-adaptive conformal calibration}. At test time, our method can efficiently estimate uncertainty intervals with pixel-marginal coverage guarantees. In addition, \ours can also predict pixel-wise conditional probability density estimates without any built-in distributional assumptions. We evaluate our method on several denoising problems, accelerated magnetic resonance imaging, and quantitative phase microscopy. Our method consistently produces tighter uncertainty intervals than prior conformal methods at the same coverage level, can predict plausible conditional distributions for different tasks, and in some cases, high-uncertainty regions can help us locate hallucinations in a model's prediction. 
    
    %We propose \textit{\ours}, a quantile uncertainty training and conformal calibration technique that enables nonlinear, spatially-adaptive scaling of quantile predictions to enable tighter uncertainty estimates for image-valued problems. Using a U-Net architecture with a quantile embedding, \ours learns the full conditional quantile distribution for each image and, after conformal calibration, can efficiently predict pixel-wise uncertainty intervals with guaranteed coverage. In addition, \ours can also predict pixel-wise conditional probability density estimate without any built-in distributional assumptions. We evaluate our method on several denoising problems, accelerated magnetic resonance imaging, and quantitative phase microscope. We show that our method consistently produces tighter uncertainty intervals that prior methods at the same coverage level. 
\end{abstract}

\section{Introduction}

Deep learning offers tremendous promise in scientific and medical imaging by enabling the algorithmic extraction of meaningful signals from sparser, noisier, and more underdetermined measurements~\citep{ongie2020deeplearningtechniquesinverse, alshardan2024deep, barbastathis2019use}. For example, deep learning has enabled image restoration in fluorescence microscopy from 60-fold fewer photons~\cite{weigert2018content}, can accelerate magnetic resonance imaging by requiring fewer k-space measurements for compelling reconstruction~\cite{zbontar2018fastmri}, and can aid in black hole recovery from sparse interferometric measurements~\cite{feng2023score}. However, this promise is counteracted by significant risk: \textit{can we trust the model's prediction? How can we determine if the model's prediction is incorrect?}

Deep learning models will often produce compelling, high-resolution reconstructions, but these predictions may not coincide with reality. Failures and hallucinations (i.e., predictions that do not coincide with reality) are incredibly hard to pinpoint. However, when using deep learning for scientific and medical imaging, it is critical for scientists and doctors to understand how and when a model's prediction may be wrong. In this paper, we introduce a method to predict both statistically valid error bars and pixel-wise conditional probability density functions for imaging inverse problems. These can alert users to regions of high predicted uncertainty where failures or hallucinations may be more likely, and allow them to probe the estimated distribution of possibilities. 

\begin{figure}
  \centering
\includegraphics[width=0.9\linewidth]{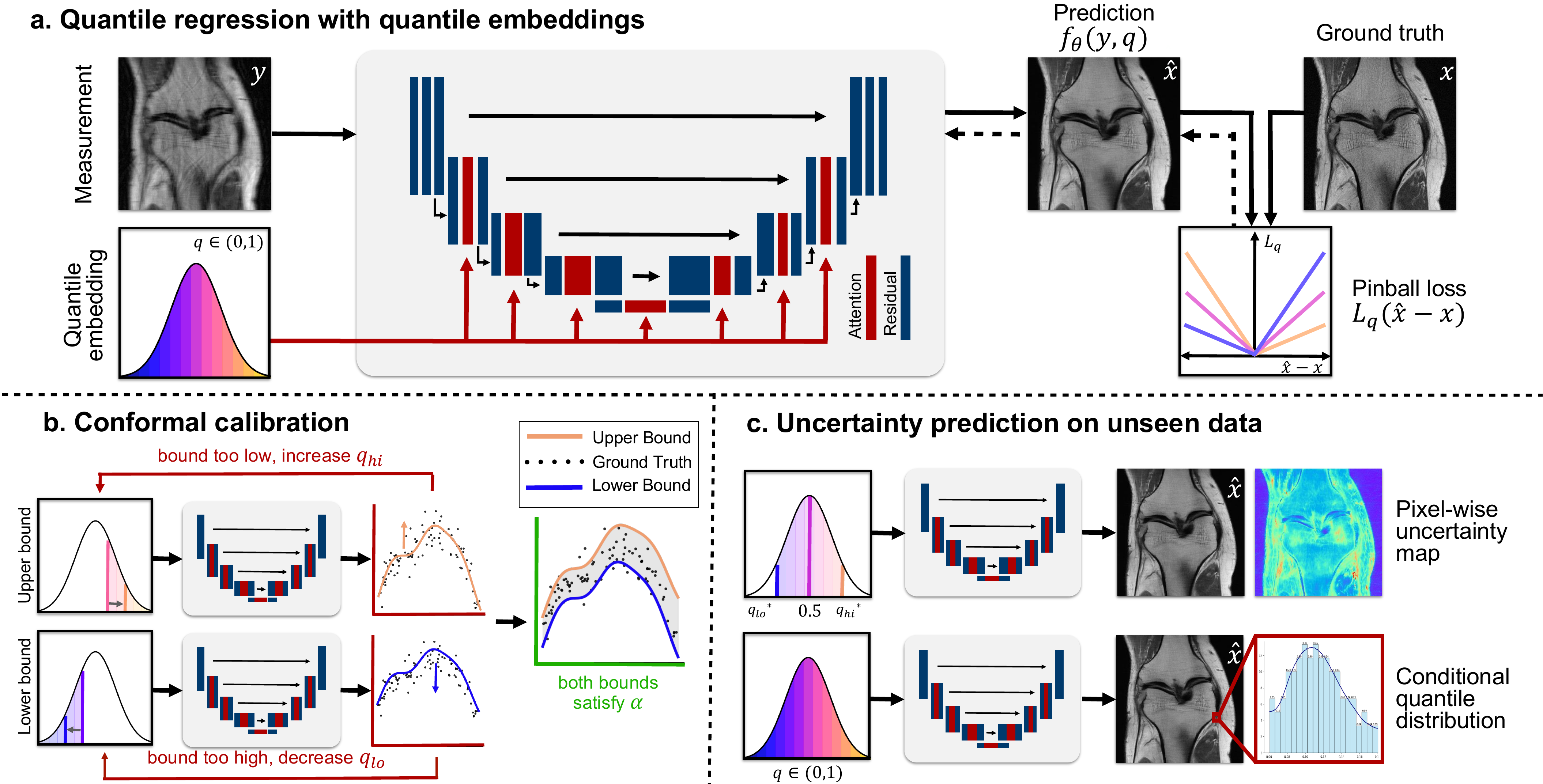}
    \caption{\textbf{\ours Overview. a) Quantile Regression with Quantile Embedding:} During training, a neural network with a quantile embedding predicts an image as a function of the measurement and quantile, $q$. The quantile embedding is randomly sampled  ($q \in (0, 1)$) and the value of $q$ determines the asymmetry of the pinball loss, enabling the model to learn a range of conditional quantiles. \textbf{b) Conformal Calibration:} During calibration, the predictive bounds ($q_{\text{lo}}, q_{\text{hi}}$) are iteratively adjusted on a held-out dataset to satisfy the desired miscoverage level $\alpha$. \textbf{c) Uncertainty Prediction on Unseen Data:} At test time, the model can be queried to predict the mean image, a pixel-wise uncertainty map, or a conditional probability density function at each pixel. 
    %At test time, the model is queried with $q_{\text{lower}}$, $q_{0.5}$, and $q_{upper}$ to produce the mean prediction and a corresponding pixel-wise uncertainty map. Querying the full range of quantile values enables the prediction of the pixel-wise distribution.
    }
  \label{fig:architecture}
\end{figure}

Our method, Quantile Uncertainty Training and Conformal Calibration (\ours)\footnote{Pronounced like cutesy, \textipa{/"kju:t.si/}},  builds on previous work in conformal prediction for image regression~\citep{angelopoulos2022image, teneggi2023trust}, but is specifically designed for high-dimensional imaging data. While prior methods use a linear, constant scaling to obtain calibrated uncertainty intervals and ignore spatial correlations in high-dimensional imaging data, our approach leverages a learned, nonlinear scaling that better captures the underlying distribution of such data. 
%Prior work on applying conformal methods for uncertainty prediction in imaging takes a heuristic uncertainty estimator (e.g., upper and lower bounds predicted by a feed-forward network or by sampling a diffusion model multiple times), and scales these heuristic uncertainty intervals with a uniform constant(s) to obtain coverage on a calibration set.
Our contributions are the following:

\begin{enumerate}
    \item We introduce a practical simultaneous quantile regression method for high-dimensional imaging inverse problems based on a U-Net with quantile conditioning. We show how this can be paired with conformal calibration, such as risk-controlling prediction sets, to produce tighter uncertainty intervals than previous conformal methods~\citep{angelopoulos2022image, teneggi2023trust}. This method is fast and practical, producing tight, calibrated uncertainty-interval predictions with only two forward passes through the network. 
    \item We introduce a method to estimate and calibrate a pixel-wise conditional probability density function ($p(\mathbf x| \mathbf y)$). By querying the same, quantile-conditioned network several times, we can obtain an estimate of the conditional probability density function. This conditional distribution does not include any Gaussian distributional assumptions, allowing us to visualize unknown pixel-wise conditional distributions. 
    \item We apply and evaluate our method on five imaging inverse problems. Two of these problems, synthetic Gaussian and Poisson denoising, allow us to probe and evaluate our method on a known noise distribution. The other three problems: microscopy denoising, compressive magnetic resonance imaging, and quantitative phase imaging, test our performance on challenging, real imaging data. Our method achieves state-of-the-art results on these tasks for uncertainty interval prediction, predicting smaller intervals than previous conformal methods, while being calibrated to the same coverage. Furthermore, we show  examples in which our prediction of a high-uncertainty region helps us identify model hallucinations.  
\end{enumerate}

\section{Background and Notation}
First, we briefly introduce the relevant background and notation. See \ref{related_work} for a comprehensive related works section. Our method is based on conformal prediction,  which constructs predictive intervals with finite-sample marginal coverage guarantees at a user-specified level ($1-\alpha$) by transforming uncertainty estimates using a held-out calibration dataset~\citep{angelopoulos2021gentle, correia2024information, romano2019conformalized}. In general, conformal methods take heuristic or model-derived uncertainty estimates and transform them into statistically valid predictive intervals using a held-out calibration dataset. The simplicity, speed, and distribution-free nature of conformal methods have made them popular across classification~\cite{angelopoulos2020uncertainty, ding2023class}, language modeling~\citep{quach2023conformal, campos2024conformal}, robotics~\citep{lindemann2023safe, lekeufack2024conformal}, and protein design~\citep{fannjiang2022conformal}. Recently, conformal prediction has been applied to image-to-image regression for pixel-wise uncertainty prediction~\citep{angelopoulos2022image, ye2025learned, teneggi2023trust}. However, existing methods, like Im2Im~\cite{angelopoulos2022image} and K-RCPS~\cite{teneggi2023trust}, use a fixed calibration procedure that is not spatially-adaptive. We describe existing methods below, and then outline how our method enables spatially-adaptive conformal calibration, resulting in smaller interval sizes for imaging problems.

\subsection{Conformal prediction for image-valued problems}
In imaging inverse problems, the goal is to predict an image, $\mathbf x \in \mathbb{R}^{d_x}$, from a measurement, $\mathbf y \in 	\mathbb{R}^{d_y}$, given some imaging operator/degradation and noise $n$, $\mathbf y=A(\mathbf x)+n$. This general framework applies to a wide range of imaging problems, including those that are ill-conditioned or underdetermined. We assume that we have access to matched input-output pairs, \( \mathcal{D}_t = \{(\mathbf x_i, \mathbf y_i)\}_{i=1}^{N_t} \), that are randomly sampled from the unknown joint distribution $p(\mathbf x, \mathbf y)$. In addition, we assume access to an underlying predictor, $f(\mathbf y)$\footnote{Note that we adopt the notation $\hat{\mathbf x}=f(\mathbf y)$, which is commonly used in the field of inverse problems instead of $\hat{\mathbf y} = f(\mathbf x)$, which is more common in the machine learning literature.}, which maps a measurement, $\mathbf y$, to an image estimate, $\hat{\mathbf x}$. 

On a high level, the goal of conformal prediction for image regression problems~\cite{angelopoulos2022image, teneggi2023trust} is to predict an uncertainty interval for each pixel in the image $\hat{\mathbf x}$ that contains the true pixel values with a user-specified probability. This uncertainty interval $[\hat{\mathbf x} - \hat{l}(\mathbf y), \hat{\mathbf x} + \hat{u}(\mathbf y)]$ will have a width of $\hat{l}(\mathbf y)$ in the lower direction, and a width of $\hat{u}(\mathbf y)$ in the upper direction. But how do we practically obtain these intervals? Prior work has shown that pixel-wise quantile regression offers state-of-the-art performance and is practical to implement~\cite{angelopoulos2022image}. In quantile regression, a neural network is trained to estimate a conditional quantile instead of a mean estimate. This can be accomplished using an asymmetric pinball loss:
\begin{equation}
    L_{q}( x, \hat{ x}) = \begin{cases}
q \cdot \lvert  x - \hat{ x} \rvert & \text{if }  x -\hat{x} \geq  0 \\
(1-q) \cdot \lvert  x - \hat{x} \rvert & \text{otherwise}.
\end{cases}
\label{eq:pinball_loss}
\end{equation}
where $\hat{x}$ denotes the predicted value, $x$ represents the ground truth, and $q \in (0, 1)$ is the quantile of interest. When $q > 0.5$, the loss assigns a greater penalty to underestimations (i.e., $\hat{x} < x$), encouraging the model to predict higher values. Conversely, when $q < 0.5$, overestimations incur a larger penalty, biasing predictions downward. This asymmetry enables the model to learn conditional quantiles of the target distribution, in contrast to losses like mean squared error (MSE), which are symmetric and designed to estimate the conditional mean. To train a predictor for the 90\% uncertainty interval, the lower and upper bounds can be set to estimate the 0.95 and 0.05 quantiles, respectively. Training a predictor with quantile loss results in a heuristic upper and lower interval prediction, $\mathcal{C}(\mathbf y)= [\hat{\mathbf x} - \tilde {l}(\mathbf y), \hat{\mathbf x} + \tilde{u}(\mathbf y)]$.  However, this interval does not necessarily contain the ground-truth pixel values with the desired probability. This can be remedied through conformal calibration, in which these heuristic bounds are scaled using a held-out calibration dataset, $\mathcal{D}_c=\{(\mathbf{x}_i, \mathbf{y}_i)\}_{i=1}^{N_c}$, until they contain the desired fraction of ground truth pixels. %A common procedure for this is risk controlling prediction sets (RCPS)~\cite{bates2021distribution}.
%See [Appendix] for a summary of the RCPS procedure, including key assumptions and guarantees. 

Im2Im uses the Risk Controlling Prediction Sets (RCPS) procedure~\cite{bates2021distribution}, which determines a scaling factor $\hat \lambda$ such that the calibrated prediction intervals, $ C_\lambda (\mathbf y) = [ \hat{\mathbf x} - \hat \lambda \tilde{l}(\mathbf y), \hat{\mathbf x} + \hat \lambda \tilde{u}(\mathbf y)]$, achieve the desired coverage $(1 - \alpha)$ with high probability.
%At a high-level, image-to-image regression methods that leverage RCPS will scale the lower and upper bounds of the interval by a scaling factor, $\hat \lambda$, to obtain the calibrated prediction interval, $ C_\lambda = [ \hat{\mathbf x} - \hat \lambda \tilde{l}(\mathbf y), \hat{\mathbf x} + \hat \lambda \tilde{u}(\mathbf y)]$
%final interval widths:  $ \hat{l}(\mathbf y) = \hat \lambda \tilde{l}(\mathbf y)$, $\hat{u}(\mathbf y) = \hat \lambda \tilde{u}(\mathbf y)$.
%\textbf{Cassandra:} To provide uncertainty bounds with statistical guarantees, the predicted upper and lower bounds are calibrated using a held-out calibration dataset $\mathcal{D}_c=\{(\mathbf{x}_i, \mathbf{y}_i)\}_{i=1}^{N_c}$. 
RCPS frames calibration as a risk-control problem. The goal is to choose $\lambda$ so that the population miscoverage risk $R(\lambda) = \mathbb{P}(\mathbf{X} \notin C_{\lambda}(\mathbf{Y}))$ stays below a user-chosen miscoverage level $\alpha$. Since $R(\lambda)$ is unknown, we estimate it from the calibration set via the empirical risk:
\[
\hat{R}(\lambda) = \frac{1}{N_c} \sum_{i=1}^{N_c} \mathbbm{1}\{\mathbf{x}_i \notin C_{\lambda}(\mathbf{y}_i)\}.
\] 
Because $\hat{R}(\lambda)$ is a finite-sample estimate of the population risk $R(\lambda)$, RCPS controls $R(\lambda)$ via a pointwise upper confidence bound $\hat{R}^+(\lambda)$ at confidence level $1-\delta$ (e.g., via the Hoeffding--Bentkus inequality). The RCPS scaling parameter is then chosen as
\[
\hat{\lambda} = \inf \left\{ \lambda : \hat{R}^+(\lambda') \leq \alpha \;\; \text{for all } \lambda' \geq \lambda \right\},
\]
where $\alpha \in (0,1)$. Since the prediction intervals are nested (larger $\lambda$ yields wider, more conservative intervals), this selection guarantees $\mathbb{P}\bigl(R(\hat{\lambda}) \leq \alpha\bigr) \geq 1 - \delta.$ At inference time, the single calibrated constant $\hat{\lambda}$ is applied uniformly to both the predicted upper and lower bounds for every new measurement $\mathbf{y}$, producing the final prediction interval $C_{\hat{\lambda}}(\mathbf{y})$. 

%Letting $\hat{u}(\mathbf{y})$ and $\hat{l}(\mathbf{y})$ denote the model's raw upper and lower bounds (as introduced in Section~3.1), this scaling takes the form
%\[
%C_{\hat{\lambda}}(\mathbf{y}) = \bigl[\,\hat{l}(\mathbf{y}) - \hat{\lambda}\,(\hat{u}(\mathbf{y}) - \hat{l}(\mathbf{y})),\ \hat{u}(\mathbf{y}) + \hat{\lambda}\,(\hat{u}(\mathbf{y}) - \hat{l}(\mathbf{y}))\,\bigr].
%\]

\subsection{Non-spatially adaptive conformal calibration}
\begin{figure}
  \centering
  \includegraphics[width=\linewidth]{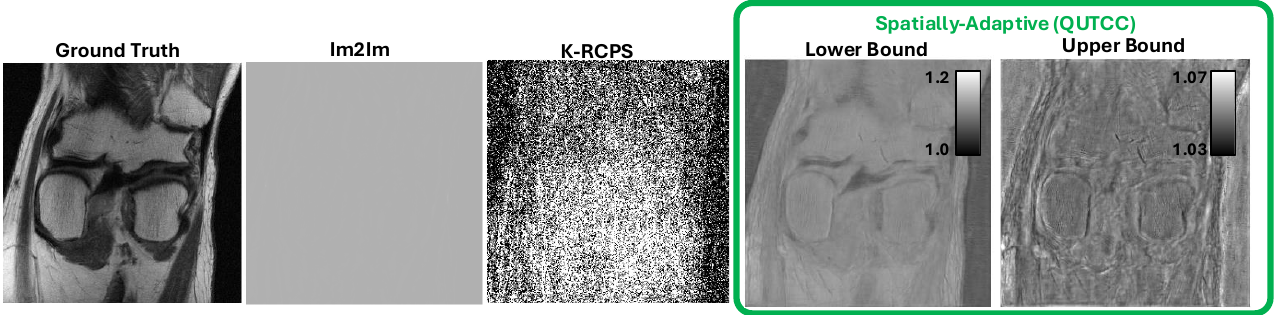}
  \caption{\textbf{Scaling parameter visualization}: Im2Im uses a single global scaling parameter, $\lambda$, $K$-RCPS calibrates $K$ group-wise parameters $\lambda_k$ over a fixed pixel partition; QUTCC produces an implicit pixel-wise scaling via direct quantile calibration for tighter, more adaptive bounds.
 }  
  \label{sup:diff_lambdas}
\end{figure}

While RCPS provides rigorous coverage guarantees, the calibrated parameter, $\hat \lambda$, acts as a uniform scalar applied identically to every pixel, regardless of local image content. This limitation was partially addressed by $K$-RCPS~\cite{teneggi2023trust}, which partitions the image into $K$ groups via a fixed partition matrix (e.g. based on Otsu's method~\cite{otsu1979threshold}, or another heuristic) and assigns a separate scaling parameter $\hat \lambda_k$ to each group. Therefore, certain image regions can have different scaling parameters, yielding smaller uncertainty intervals compared to Im2Im at the same coverage level. However, the partition is fixed across all images and chosen independently of any individual measurement $\mathbf{y}$. An example of this global and K-partition scaling is shown in Fig.~\ref{sup:diff_lambdas} for an MRI measurement. Neither method can effectively adapt its scaling to features within an image: high and low uncertainty regions within an image are calibrated the same way. 
%As a result, neither method is \emph{spatially adaptive} in a meaningful sense: high and low uncertainty regions within a single image cannot be calibrated differently. A further limitation is that both procedures couple the upper and lower bounds through a single scaling factor. When the heuristic predictor is asymmetrically miscalibrated, for example, when $\hat{u}(\mathbf{y})$ is reliable but $\hat{\ell}(\mathbf{y})$ systematically misses- $\lambda$ must be inflated to cover the worse bound, needlessly widening the well-calibrated one. An example of the calibrated $\hat \lambda$ for RCPS, and K-RCPS is shown in Fig.~\ref{sup:diff_lambdas}. 

% During conformal calibration, we use a small calibration dataset, \( \mathcal{D}_c = \{(\mathbf x_i, \mathbf y_i)\}_{i=1}^{N_c} \), to search over different values of the input parameters $q_{hi}$ and $q_{lo}$ until we reach the desired coverage. That is, the calibrated constructed interval, $\mathcal{C}\in [f_\theta(\mathbf y, q_{\text{lo}}^*), f_\theta(\mathbf y, q_{\text{hi}}^*)]$, contains at least $1-\alpha$ of the ground truth pixels ~\citep{angelopoulos2022conformal}. Formally, this means that:

% \begin{equation}
%     E[\mathbf x_{\text{test}} \in C(\mathbf y_{\text{test}}) ] \geq 1-\alpha 
% \label{eq:coverage}
% \end{equation}
% \textcolor{red}{
% \begin{equation}
% \mathbb{E} \Big[ \mathbf{x}_{\text{test}}[k] \in C(\mathbf{y}_{\text{test}}[k]) \Big] \ge 1 - \alpha, 
% \quad \forall k \in \{1, \dots, K\},
% \label{eq:coverage}
% \end{equation}}

% \textcolor{green}{
% \begin{equation}
% % \hat{R}\!\left(q_{\text{lower}}, q_{\text{upper}}\right)
% % \;=\;
% % \frac{1}{K} \sum_{k=1}^{K}
% % \mathbb{E}\!\left[
% % \mathbb{I}\!\left\{
% % \mathbf{x}_k
% % \notin
% % C\!\left(\mathbf{y}_k\right)
% % \right\}
% % \right]
% % \;\leq\; \alpha.
% \hat{R}(q_{\text{lo}^*}, q_{\text{hi}^*}) = \frac{1}{K}\sum_{k=1}^{K} \mathbb{E}\!\left[ \mathbb{I}\left\{ \mathbf{x}_k \notin C(\mathbf{y}_k) \right\} \right] \le \alpha.
% \label{eq:coverage}
% \end{equation}
% % }

\section{Methodology}
What if we could predict how uncertainty varies across the image to determine continuous, spatially-adaptive scaling parameters? Rather than calibrating a global scalar parameter~\cite{angelopoulos2020uncertainty}, or K-parameters~\cite{teneggi2023trust} based on a pre-computed heuristic, we propose to calibrate the prediction interval based on a conditioning input to a neural network in which a neural network is conditioned on the upper and lower quantiles, $\hat q_{\text{hi}}, \hat q_{\text{lo}}$. This interval,

\begin{equation}
    \mathcal{C}_{q_{\text{lo}}, q_{\text{hi}}}\in [f_\theta(\mathbf y, \hat q_{\text{lo}}), f_\theta(\mathbf y, \hat q_{\text{hi}})],
\end{equation}

is a function of a neural network, and enables continuous, \textit{spatially-adaptive calibration} that leverages information across an image to determine how to best scale different image regions.

We obtain a heuristic uncertainty interval by setting the network's conditioning parameters to high and low quantile values, e.g., $\tilde q_{\text{hi}}=0.95$, $\tilde q_{\text{lo}}=0.05$. During conformal calibration with RCPS, these parameters $\hat q_{hi}$ and $\hat q_{lo}$ are updated until the desired coverage is reached on the calibration dataset (e.g. $\hat q_{\text{hi}}=0.97$, $\hat q_{\text{lo}}=0.06$). While we only calibrate these two input parameters, this results in a spatially-varying scaling across the entire image. This scaling is non-linear, spatially-varying, and learned from the training dataset. An example of our learned pixel-wise scaling is shown in Fig.~\ref{sup:diff_lambdas}(right), with full details on how we calculate this effective image-space scaling in~\ref{sec:adaptive_scaling_vs_krcps}.  %, effectively using information across the entire image to scale different regions of the image during conformal calibration. 

After calibration, the network can provide a pixel-wise uncertainty map for new, unseen data \textit{or} can be queried to predict a conditional distribution for any pixel in the image. Our method, \ours, is summarized in Fig.~\ref{fig:architecture}. We elaborate on how we train our network through simultaneous quantile regression, how we calibrate the heuristic intervals through conformal calibration, and how we use the network to predict a conditional distribution in the sections below.
%Section about Model Architecture + Attention

\subsection{Simultaneous quantile regression}
%\subsection{Network Architecture}
We leverage ideas from simultaneous quantile regression to train a single neural network, $f_{\theta}(\mathbf y, q)$, to predict any conditional quantile, $q$, of the joint distribution. The network $f_{\theta}(\mathbf y, q)$ -- an attention U-Net~\citep{ronneberger2015u,oktay2018attention} with a quantile-embedding -- is conditioned on the parameter $q$, to predict the conditional quantile. The image estimate is obtained by querying the network at $q=0.5$, which gives us the median image: 

\begin{equation}
    \hat{\mathbf x} = f_\theta(\mathbf y, q=0.5).
\end{equation}

To train the neural network to predict an arbitrary conditional quantile image, we use pinball loss, (Eq.~\ref{eq:pinball_loss}). At each training step, the quantile parameter, $q$ is randomly sampled and used to both condition the network \textit{and} as an input to the loss function. This allows the model to learn the full conditional quantile function, rather than a discrete, fixed quantile value as in prior image-to-image regression methods~\citep{angelopoulos2022image, ye2025learned}. The total loss is given by:
% \begin{equation}
% \mathcal{L}_{\text{total}}(\theta) =  \sum_{i =1}^{N_t} \mathcal{L}_{q}(x_i, f_{\theta}(y_i, q)),
% \label{eq:total_loss}
\begin{equation}
\mathcal{L}_{\text{total}}(\theta) = \mathbb{E}_{[(\mathbf x, \mathbf y) \sim D_{\text{t}}, {q \sim \mathcal{U}(0,1)]}} \Big[ \mathcal{L}_{q}\big(\mathbf x, f_{\theta}(\mathbf y, q)\big) \Big],
\label{eq:total_loss_full_expectation}
\end{equation}

where $f_{\theta}$ is a neural network with parameters $\theta$, and $f_{\theta}(\mathbf y_i, q)$ is the output of the neural network given an input measurement, $\mathbf y_i$ and quantile value $q$. The loss is an expectation over both the data distribution $(x, y) \sim D_{\text{t}}$ and quantile levels $q \sim \mathcal{U}(0,1)$, explicitly reflecting the uniform sampling used during training. This loss is minimized with the Adam optimizer~\citep{kingma2014adam} using backpropagation. The neural network weights are shared across different quantile predictions, limiting quantile crossing. The proposed network is shown in Fig.~\ref{fig:architecture}, and the full architecture and training details are described in Appendix~\ref{sup:experiment_details},~\ref{sup:model_arch}.  %During training, the neural network learns a full conditional quantile function, rather than a discrete, fixed quantile value as in prior image-to-image regression methods~\citep{angelopoulos2022image, ye2025learned}. 

%Section about how we calibrate the bounds
\subsection{Conformal calibration}
\label{Calibration_step}
After the neural network is trained to obtain heuristic uncertainty interval predictions, we use an RCPS-style calibration to ensure statistical coverage. Rather than calibrating a scaling parameter that directly scales the predicted bounds, we find the neural network conditioning parameters ($q_{lo}$, $q_{hi}$) that achieve the target coverage, to obtain the final calibrated parameters (${\hat q_{lo}}$, ${\hat q_{hi}}$). We calibrate the lower and upper quantile bounds independently to provide marginal pixelwise coverage across the calibration dataset. To satisfy a target total miscoverage rate of $\alpha$, the calibration process allocates half of this error budget to each bound, so that the lower and upper bounds each capture violations at a marginal rate no greater than $\alpha/2$. Thus, if $\mathbf x_i[k]$ denotes the ground truth at pixel $k$ of image $i$ and $[f_\theta(\mathbf y_i, \hat q_{\text{lo}})[k], f_\theta(\mathbf y_i, \hat q_{\text{hi}})[k]]$ is the calibrated predicted interval, we enforce: 
\begin{equation}
\frac{1}{N_c K} \sum_{i=1}^{N_c} \sum_{k=1}^{K}
\mathbbm{1}\!\left\{ x_i[k] < f_\theta(\mathbf y_i, \hat q_{\text{lo}})[k] \right\}
\;\leq\; \frac{\alpha}{2},
\quad
\frac{1}{N_c K} \sum_{i=1}^{N_c} \sum_{k=1}^{K}
\mathbbm{1}\!\left\{ x_i[k] >f_\theta(\mathbf y_i, \hat q_{\text{hi}})[k] \right\}
\;\leq\; \frac{\alpha}{2}.
\end{equation}
% \begin{equation}
% \frac{1}{N_c K} \sum_{i=1}^{N_c} \sum_{k=1}^{K}
% \mathbbm{1}\!\left\{ x_i(k) < l_{q_{\text{lo}}^*}(y_i)(k) \right\}
% \;\leq\; \frac{\alpha}{2},
% \quad
% \frac{1}{N_c K} \sum_{i=1}^{N_c} \sum_{k=1}^{K}
% \mathbbm{1}\!\left\{ x_i(k) > u_{q_{\text{hi}}^*}(y_i)(k) \right\}
% \;\leq\; \frac{\alpha}{2}.
% \end{equation}
By a union bound, these two conditions together guarantee a total marginal miscoverage rate of at most $\alpha$.
Pseudocode for this calibration process is provided in  \textbf{Algorithm 1}. At each step, we compute the marginal miscoverage from the quantile upper and lower bounds over the entire calibration dataset. If the violation rate for a bound exceeds $\alpha/2$, we relax the corresponding bound to widen the interval; otherwise, we tighten it. This process proceeds via a binary search over the quantile space until the desired coverage is reached. Conducting a binary search over the quantile space assumes that the learned quantile function is monotonic. In practice, this function is mostly monotonic with a few rare violations, which occur in background regions (See Appx Tbl.~\ref{tbl:quantile_crossing}). Note that we adjust the per-bound error rate, $\alpha' \gets \frac{\alpha}{2} - \frac{1-\alpha/2}{N_c}$, to account for the finite calibration dataset size~\citep{angelopoulos2021gentle, vovk2012conditional}. At the end of this procedure, we can obtain a constructed interval $\mathcal{C} = [f_\theta(\mathbf y, \hat q_{\text{lo}}), f_\theta(\mathbf y, \hat q_{\text{hi}})]$ that controls the risk.

\begin{algorithm}[H]
\caption{Calibrating Quantile Bounds}
\label{alg:calibrating_quantile_bounds}
\begin{algorithmic}[1]
\REQUIRE Calibration risks $R_{\text{lower}}(\cdot)$, $R_{\text{upper}}(\cdot)$;
         target level $\alpha$; calibration size $N_c$; tolerance $\epsilon$
\STATE Compute adjusted error: $\alpha' \gets \alpha - \frac{1-\alpha}{N_c}$
\STATE Set per-bound budget: $\alpha'' \gets \alpha'/2$
\STATE $\hat q_{\text{lo}} \gets$ largest $q \in [0,1]$ with $R_{\text{lower}}(q) \le \alpha''$ \hfill (via binary search to tolerance $\epsilon$)
\STATE $\hat q_{\text{hi}} \gets$ smallest $q \in [0,1]$ with $R_{\text{upper}}(q) \le \alpha''$ \hfill (via binary search to tolerance $\epsilon$)
\STATE \textbf{return} $(\hat q_{\text{lo}},\, \hat q_{\text{hi}})$
\end{algorithmic}
\end{algorithm}

% Pseudocode for this calibration process is provided in  \textbf{Algorithm 1}. At each step, we compute the miscoverage from the quantile upper and lower bounds over the entire calibration dataset. If the violation rate for a bound exceeds the adjusted $\alpha$, we relax the corresponding bound; otherwise, we tighten it. This process proceeds via a binary search over the quantile space until the desired coverage is reached. Conducting a binary search over the quantile space assumes that the learned quantile function is monotonic. In practice, this function is mostly monotonic with a few rare violations, which occur in background regions (See Appx Tbl.~\ref{tbl:quantile_crossing}). Note that we adjust the error rate, $\alpha' \gets \alpha - \frac{1-\alpha}{N_c}$ to account for the finite calibration dataset size~\citep{angelopoulos2021gentle, vovk2012conditional}. At the end of this procedure, we can obtain a constructed interval $\mathcal{C}\in [f_\theta(\mathbf y, q_{\text{lo}}^*), f_\theta(\mathbf y, q_{\text{hi}}^*)]$ that satisfies Eq.~\ref{eq:coverage}.

\subsection{Estimating the conditional distribution}
Our approach offers an additional benefit: since it is trained to predict the full quantile function rather than a single fixed quantile, we can recover an estimate of the entire quantile function at each pixel. This is accomplished by querying the network $\hat{f}(\mathbf y,q)$ over a range of quantile levels $q \in (0, 1)$, from which we construct a pixel-wise conditional PDF that serves as a standalone tool for predicting the underlying conditional distribution, $p( x | \mathbf y)$ for each pixel in $\hat{\mathbf x}$ \footnote{In this section, since we predict pixel-wise PDFs, we refer to $\mathbf{x}[k]$ as $x$, the pixel value at index k}. Under standard regularity conditions—namely, that the pixel-wise distribution admits a probability density function $p_k$, the quantile function is differentiable, and the density $p_k$ is positive on its support—the following standard identity from probability theory holds:  the PDF is inversely proportional to the rate of change of the quantile function. Applying this identity, we can estimate the conditional PDF $\hat{p}_k(x| \mathbf y)$ for pixel intensity $x_j := \hat{f}(\mathbf y,q_j)$ as:
\begin{equation}
    \hat{p}_k(x_j| \mathbf y) = \left(\frac{\partial \hat{f}(\mathbf y,q_j)}{\partial q}\right)^{-1}, \quad q_j \in \{q_1, \ldots, q_n\} \subset (0,1).
\end{equation}
We obtain the derivative $\frac{\partial \hat{f}(\mathbf y,q_j)}{\partial q}$ through numerical approximation using finite differences. In practice, we query our neural network at different quantile levels $\{q_j\}_{j \in [n]}$, where $j$ represents the quantile level being queried, to obtain predictions $x_j := \hat{f}(\mathbf y,q_j)$, which in turn can be used to compute the derivative via finite differences:
\begin{equation}
\frac{\partial \hat{f}(y,q_j)}{\partial q} \approx \frac{\hat{f}(\mathbf y,q_{j+1}) - \hat{f}(\mathbf y,q_{j-1})}{q_{j+1} - q_{j-1}}.
\end{equation}
We finally obtain estimates of the PDF at points $x_1, \cdots, x_n$ by taking the inverse of $\frac{\partial \hat{f}(\mathbf y,q_j)}{\partial q}$ for all $j \in [n]$. While this provides a useful estimate of the underlying distribution, we can further enhance these estimates with statistical guarantees through conformal calibration. To construct a conformally calibrated PDF, we first specify the desired coverage levels for multiple quantile pairs. For example, for $90\%$ coverage we initialize with quantiles $0.05$ and $0.95$, while for $60\%$ coverage we use $0.20$ and $0.80$. These initial bounds are then calibrated using Algorithm~\ref{alg:calibrating_quantile_bounds} to produce bounds that guarantee the target coverage. By performing this calibration for multiple quantile levels $\{q_j\}_{j=1}^n$ and their associated coverage rates $\{1-\alpha_i\}_{j=1}^n$, we obtain a set of conformally calibrated predictions,
\begin{equation}
\{\hat{f}^{\mathrm{conf}}(\mathbf y, \hat q_j)\}_{j=1}^n,
\end{equation}
from which the pixel-wise conditional PDF is reconstructed via finite differences:
\begin{equation}
\hat{p}_k^{\mathrm{conf}}\!\big(\hat{f}^{\mathrm{conf}}(\mathbf y, \hat q_j ) | \mathbf y \big) 
= \left( \frac{\hat{f}^{\mathrm{conf}}(\mathbf y, \hat q_{j+1}) - \hat{f}^{\mathrm{conf}}(\mathbf{y}, \hat q_{j-1})}{\hat q_{j+1} - \hat q_{j-1}} \right)^{-1}.
\end{equation}
The resulting PDF then has the statistical guarantees of the conformal calibration procedure, providing pairwise coverage at each calibrated quantile level. Producing the pixel-wise PDF requires multiple forward passes through the network, increasing computational time. In addition, this procedure relies on numerical differentiation, which could be sensitive to the number and placement of the queried quantiles, see Fig.~\ref{sup:increasing_quantiles}.

\section{Results}
\label{sec:results}
For evaluation, we compare MC-Dropout~\cite{gal2016dropout}, Deep Ensembles,~\citep{lakshminarayanan2017simple}, Im2Im\footnote{To ensure that our performance improvements come from our uncertainty quantification technique and not network improvements, we upgrade Im2Im to use the same architecture and depth as \ours}~\citep{angelopoulos2022image}, Im2Im-Asymm\footnote{This is Im2Im with 2 calibrated lambdas- one for each predicted bound}, K-RCPS~\cite {teneggi2023trust}, and our method QUTCC on five imaging inverse problems: accelerated MRI~\citep{zbontar2018fastmri}, quantitative phase imaging (QPI)~\citep{pinkard2024berkeley}, and denoising under real-noise, synthetic Poisson, and Gaussian noise ~\citep{zhang2019poisson}. MC-Dropout and Deep Ensembles are not conformal methods and provide no formal coverage guarantees, but they yield approximate uncertainty bounds via posterior sampling. Both are computationally expensive at inference: Deep Ensembles require training $n$ separate models and performing $n$ forward passes per input, while MC-Dropout uses a single model but still requires $n$ stochastic forward passes with active dropout to estimate bounds. In contrast, the conformal methods (Im2Im, Im2Im-Asymm, K-RCPS, and QUTCC) use a single model and require only one or a few forward passes. For fair comparison, all methods are conformally calibrated to the same target risk with $\alpha = 0.1$. Full training details are provided in Appendix~\ref{sup:experiment_details}. We evaluate predicted interval lengths stratified by pixel intensity across all tasks, and show interval size as a function of noise and intensity.  Additionally, we show an example of a region with high uncertainty in which a hallucination is present, and examples of a predicted pixel-wise conditional PDFs.
\subsection{Uncertainty interval length and risk}
We compare predicted uncertainty interval lengths stratified by pixel intensity for each task in Table~\ref{tbl:model_comparison}. QUTCC consistently achieves the best or second-best interval lengths across most tasks; when it ranks second, it is typically outperformed by MC-Dropout or Deep Ensembles. However, these gains come at substantially higher computational cost: MC-Dropout incurs a 28×–47× increase in inference time (Table~\ref{tab:inference_time}), while Deep Ensembles require roughly 10× more training time (Table~\ref{tab:training_time}). Mean interval lengths and risk values are reported in Table~\ref{table:interval_length_and_risk}. \ours achieves the smallest mean interval length on every task among conformal methods, and remains competitive with the posterior-sampling baselines despite using a single model and only two forward passes to construct each bound. 
\begin{figure}[h]
  \centering
  \includegraphics[width=\linewidth]{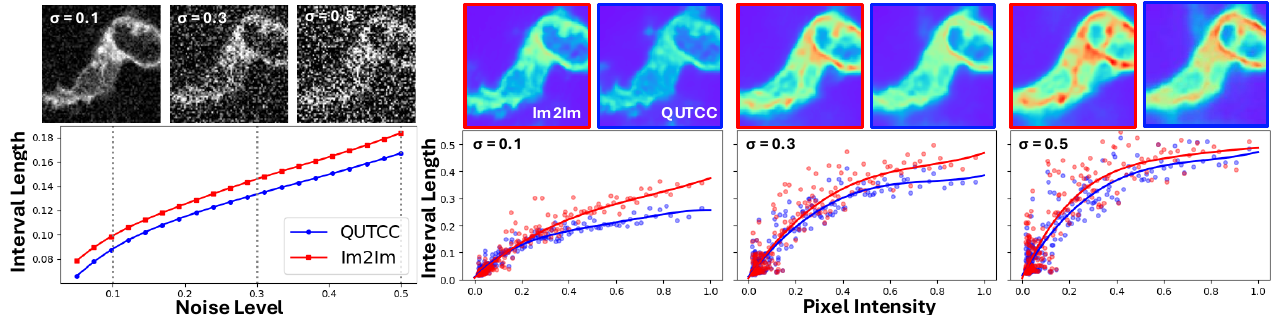}
  \caption[QUTCC uncertainty intervals under increasing Gaussian noise]{\textbf{\ours exhibits smaller uncertainty intervals in regions of high pixel intensity:} In a Gaussian denoising setting, we analyze how uncertainty interval lengths vary with pixel intensity across increasing noise levels ($\sigma = 0.1,\ 0.3,\ 0.5$). Under low-noise conditions, \ours exhibits narrower uncertainty intervals compared to the baseline. As noise increases to $\sigma = 0.5$, this advantage becomes less pronounced overall. However, \ours continues to produce shorter uncertainty intervals for high-intensity pixels (intensity $>$ 0.8), even at higher noise levels. All confidence intervals were estimated over a set of 10 samples.}
  \label{fig:increasing_noise}
\end{figure}

\definecolor{bin1}{RGB}{49,54,149}    % blue
\definecolor{bin2}{RGB}{69,117,180}   % blue-cyan
\definecolor{bin3}{RGB}{116,173,209}  % light blue
\definecolor{bin4}{RGB}{171,217,233}  % very light
\definecolor{bin5}{RGB}{255,255,191}  % yellow
\definecolor{bin6}{RGB}{253,174,97}   % orange
\definecolor{bin7}{RGB}{244,109,67}   % red-orange
\definecolor{bin8}{RGB}{215,48,39}    % red

\begin{table}[t]
\centering
\caption{Uncertainty interval length stratified by pixel intensity. Blue entries denote the smallest uncertainty intervals within each bin, while bold black entries indicate the second smallest.}
\scriptsize
\setlength{\tabcolsep}{0.3pt}
\begin{tabular*}{\textwidth}{@{\extracolsep{\fill}}llcccccc@{}}
\toprule
\multirow{2}{*}{\textbf{Task}} & \multirow{2}{*}{\textbf{Interval}} & \multicolumn{2}{c}{\textbf{Posterior-Based}} & \multicolumn{4}{c}{\textbf{Quantile-Based}} \\
\cmidrule(lr){3-4} \cmidrule(lr){5-8}
& & \textbf{MC-Drop.} & \textbf{Deep Ens.} & \textbf{Im2Im} & \textbf{Im2Im-Asym.} & \textbf{K-RCPS} & \textbf{QUTCC} \\
\midrule

\multirow{5}{*}{\textbf{Poisson}}
& \cellcolor{bin1}\textcolor{white}{0.00--0.20}& $0.123 \pm 0.107$ & $\mathbf{0.037 \pm 0.036}$ & $0.040 \pm 0.013$ & $0.039 \pm 0.013$ & $0.041 \pm 0.011$ & {\color{blue}$0.035 \pm 0.011$}\\

& \cellcolor{bin3}\textcolor{white}{0.20--0.40}& $0.371 \pm 0.148$ & $0.104 \pm 0.097$ & $0.112 \pm 0.018$ & $0.109 \pm 0.018$ & $\mathbf{0.100 \pm 0.015}$ & {\color{blue}$0.090 \pm 0.015$}\\

& \cellcolor{bin5}\textcolor{black}{0.40--0.60}& $0.643 \pm 0.213$ & $0.240 \pm 0.309$ & $0.211 \pm 0.026$ & $0.206 \pm 0.026$ & $\mathbf{0.181 \pm 0.021}$ & {\color{blue}$0.160 \pm 0.022$}\\

& \cellcolor{bin6}\textcolor{white}{0.60--0.80}& $0.663 \pm 0.215$ & $0.306 \pm 0.438$ & $0.244 \pm 0.030$ & $0.238 \pm 0.030$ & $\mathbf{0.207 \pm 0.024}$ & {\color{blue}$0.158 \pm 0.028$}\\

& \cellcolor{bin8}\textcolor{white}{0.80--1.00}& $0.771 \pm 0.250$ & $0.653 \pm 0.933$ & $0.302 \pm 0.042$ & $0.294 \pm 0.042$ & $\mathbf{0.225 \pm 0.028}$ & {\color{blue}$0.187 \pm 0.034$}\\

\midrule
\multirow{5}{*}{\textbf{Gaussian}}
& \cellcolor{bin1}\textcolor{white}{0.00--0.20}& $0.073 \pm 0.070$ & {\color{blue}$0.050 \pm 0.019$} & $0.056 \pm 0.020$ & $0.055 \pm 0.020$ & $\mathbf{0.054 \pm 0.019}$ & $0.057 \pm 0.022$\\
& \cellcolor{bin3}\textcolor{white}{0.20--0.40}& $0.198 \pm 0.121$ & {\color{blue}$0.121 \pm 0.027$} & $0.135 \pm 0.027$ & $0.132 \pm 0.026$ & $\mathbf{0.126 \pm 0.025}$ & $0.131 \pm 0.030$\\
& \cellcolor{bin5}\textcolor{black}{0.40--0.60}& $0.318 \pm 0.169$ & $0.216 \pm 0.043$ & $0.227 \pm 0.041$ & $0.222 \pm 0.040$ & $\mathbf{0.211 \pm 0.037}$ & {\color{blue}$0.206 \pm 0.046$}\\
& \cellcolor{bin6}\textcolor{white}{0.60--0.80}& $0.318 \pm 0.129$ & $\mathbf{0.198 \pm 0.050}$ & $0.239 \pm 0.040$ & $0.234 \pm 0.039$ & $0.222 \pm 0.036$ & {\color{blue}$0.190 \pm 0.048$}\\
& \cellcolor{bin8}\textcolor{white}{0.80--1.00}& $0.355 \pm 0.126$ & $\mathbf{0.230 \pm 0.061}$ & $0.276 \pm 0.045$ & $0.270 \pm 0.044$ & $0.246 \pm 0.038$ & {\color{blue}$0.205 \pm 0.051$}\\

\midrule
\multirow{5}{*}{\textbf{Real-Noise}}
& \cellcolor{bin1}\textcolor{white}{0.00--0.20}& $\mathbf{0.025 \pm 0.015}$ & {\color{blue}$0.021 \pm 0.006$} & $0.030 \pm 0.013$ & $0.030 \pm 0.013$ & $0.031 \pm 0.012$ & $0.033 \pm 0.012$\\

& \cellcolor{bin3}\textcolor{white}{0.20--0.40}& $\mathbf{0.104 \pm 0.035}$ & {\color{blue}$0.088 \pm 0.013$} & $0.166 \pm 0.017$ & $0.165 \pm 0.017$ & $0.151 \pm 0.015$ & $0.140 \pm 0.016$\\

& \cellcolor{bin5}\textcolor{black}{0.40--0.60}& $\mathbf{0.189 \pm 0.047}$ & {\color{blue}$0.152 \pm 0.021$} & $0.232 \pm 0.018$ & $0.230 \pm 0.018$ & $0.208 \pm 0.016$ & $0.202 \pm 0.018$\\

& \cellcolor{bin6}\textcolor{white}{0.60--0.80}& $0.273 \pm 0.058$ & ${\color{blue}0.233 \pm 0.037}$ & $0.266 \pm 0.017$ & $0.263 \pm 0.017$ & $\mathbf{0.236 \pm 0.015}$ & $0.248 \pm 0.021$\\

& \cellcolor{bin8}\textcolor{white}{0.80--1.00}& $0.342 \pm 0.055$ & $0.318 \pm 0.063$ & $0.278 \pm 0.014$ & $\mathbf{0.274 \pm 0.014}$ & {\color{blue}$0.197 \pm 0.014$} & $0.286 \pm 0.025$\\

\midrule
\multirow{5}{*}{\textbf{MRI}}
& \cellcolor{bin1}\textcolor{white}{0.00--0.20}& $0.132 \pm 0.051$ & {\color{blue}$0.088 \pm 0.026$} & $\mathbf{0.092 \pm 0.025}$ & $0.092 \pm 0.025$ & $0.091 \pm 0.025$ & $0.092 \pm 0.026$\\

& \cellcolor{bin3}\textcolor{white}{0.20--0.40}& $0.206 \pm 0.073$ & {\color{blue}$0.127 \pm 0.031$} & $0.132 \pm 0.029$ & $0.132 \pm 0.029$ & $0.130 \pm 0.028$ & $\mathbf{0.130 \pm 0.031}$\\
& \cellcolor{bin5}\textcolor{black}{0.40--0.60}& $0.304 \pm 0.093$ & $\mathbf{0.132 \pm 0.029}$ & $0.135 \pm 0.028$ & $0.135 \pm 0.028$ & $0.133 \pm 0.027$ & {\color{blue}$0.130 \pm 0.030$}\\
& \cellcolor{bin6}\textcolor{white}{0.60--0.80}& $0.454 \pm 0.131$ & $\mathbf{0.138 \pm 0.026}$ & $0.141 \pm 0.024$ & $0.140 \pm 0.024$ & $0.139 \pm 0.024$ & {\color{blue}$0.133 \pm 0.027$}\\
& \cellcolor{bin8}\textcolor{white}{0.80--1.00}& $0.680 \pm 0.223$ & $\mathbf{0.140 \pm 0.020}$ & $0.140 \pm 0.019$ & $0.140 \pm 0.019$ & $0.138 \pm 0.018$ & {\color{blue}$0.136 \pm 0.022$}\\

\midrule
\multirow{5}{*}{\textbf{QPI}}
& \cellcolor{bin1}\textcolor{white}{0.00--0.20}& $0.055 \pm 0.012$ & {\color{blue}$0.037 \pm 0.005$} & $0.059 \pm 0.005$ & $0.058 \pm 0.004$ & $0.061 \pm 0.005$ & $\mathbf{0.057 \pm 0.005}$\\

& \cellcolor{bin3}\textcolor{white}{0.20--0.40}& $0.059 \pm 0.144$ & {\color{blue}$0.042 \pm 0.006$} & $0.067 \pm 0.005$ & $0.066 \pm 0.005$ & $0.067 \pm 0.006$ & $\mathbf{0.063 \pm 0.006}$\\

& \cellcolor{bin5}\textcolor{black}{0.40--0.60}& $0.088 \pm 0.018$ & {\color{blue}$0.054 \pm 0.006$} & $0.085 \pm 0.006$ & $0.083 \pm 0.006$ & $0.088 \pm 0.006$ & $\mathbf{0.081 \pm 0.006}$\\

& \cellcolor{bin6}\textcolor{white}{0.60--0.80}& $0.105 \pm 0.017$ & {\color{blue}$0.059 \pm 0.006$} & $0.093 \pm 0.006$ & $0.091 \pm 0.006$ & $0.096 \pm 0.006$ & $\mathbf{0.090 \pm 0.006}$\\
& \cellcolor{bin8}\textcolor{white}{0.80--1.00}& $0.116 \pm 0.016$ & {\color{blue}$0.054 \pm 0.006$} & $0.087 \pm 0.007$ & $\mathbf{0.085 \pm 0.007}$ & $0.091 \pm 0.006$ & $0.087 \pm 0.007$\\
\bottomrule
\end{tabular*}
\label{tbl:model_comparison}
\end{table}

Interestingly, QUTCC’s performance gains lie largely in high intensity pixels. For low-intensity background pixels, where the images contain minimal signal, QUTCC exhibits performance comparable to the baselines, reflecting high model confidence in these regions. In contrast, in high-intensity, signal-rich regions corresponding to biological structure, QUTCC produces narrower prediction intervals. In Figure~\ref{fig:increasing_noise}, we compare \ours with Im2Im on Gaussian denoising under progressively increasing noise, reporting average uncertainty interval lengths as a function of the noise level. \ours consistently has smaller interval lengths at differet noise levels. Additionally, at $\sigma \in {0.1, 0.3, 0.5}$, we examine a localized region containing signal and observe that \ours yields narrower uncertainty intervals in high-intensity pixels. This advantage becomes less pronounced as noise increases. 

\begin{wraptable}[7]{r}{0.48\textwidth}
\vspace{-4pt}
\centering
\small
\setlength{\tabcolsep}{3pt}
\caption{Interval length across MRI accelerations.}
\label{tab:acceleration}
\begin{tabular}{lcccc}
\toprule
 & $4\times$ & $8\times$ & $12\times$ & $16\times$ \\
\midrule
Im2Im & 0.1121 & 0.1371 & 0.1530 & 0.1690 \\
QUTCC & 0.1116 & 0.1348 & 0.1511 & 0.1643 \\
QUTCC gain (\%) & 0.45 & 1.68 & 1.24 & 2.78 \\
\bottomrule
\end{tabular}
\vspace{-6pt}
\end{wraptable}
When is QUTCC more beneficial than scalar alternatives such as Im2Im and K-RCPS? QUTCC's advantage over global or partition-based scalar calibration depends on how much the true per-pixel uncertainty intervals vary across the image, i.e., the degree of heteroscedasticity in the ground-truth conditional distribution. When the ideal interval width is roughly uniform, a single global scalar, $\lambda$, (Im2Im) or a small set of partition-wise scalars, $\lambda_k$ (K-RCPS), are sufficient. However, as the uncertainty becomes more spatially variable, scalar calibration becomes increasingly wasteful since it must inflate intervals everywhere to cover the most uncertain regions. For example, Poisson noise is signal-dependent with higher variance at larger intensities, so the optimal interval width differs between dark and bright regions. QUTCC has higher gains in this regime (Table~\ref{tbl:model_comparison}). Additive Gaussian noise is approximately homoscedastic by construction, so a global scalar is already close to optimal and the gain is moderate. In addition, we note that this gain can grow for more ill-conditioned problems. We trained Im2Im and QUTCC across four MRI acceleration factors, Table~\ref{tab:acceleration}. As the reconstruction problem becomes more ill-conditioned at the higher acceleration factors, QUTCC's gain increases.

% As an example, we trained Im2Im and QUTCC across four MRI acceleration factors to examine whether QUTCC's benefit grows as the reconstruction problem becomes more ill-posed (Table~\ref{tab:acceleration}). QUTCC's gains generally increase with the acceleration factor, from a 0.45\% reduction in interval length at $4\times$ to 2.78\% at $16\times$. 
% This trend supports our hypothesis that higher acceleration induces more spatially varying uncertainty, giving QUTCC's spatially adaptive calibration more structure to exploit than Im2Im's single global scalar. 

\subsection{Uncertainty Visualizations}
\begin{figure*}[h]
  \centering
  \includegraphics[width=0.7\linewidth]{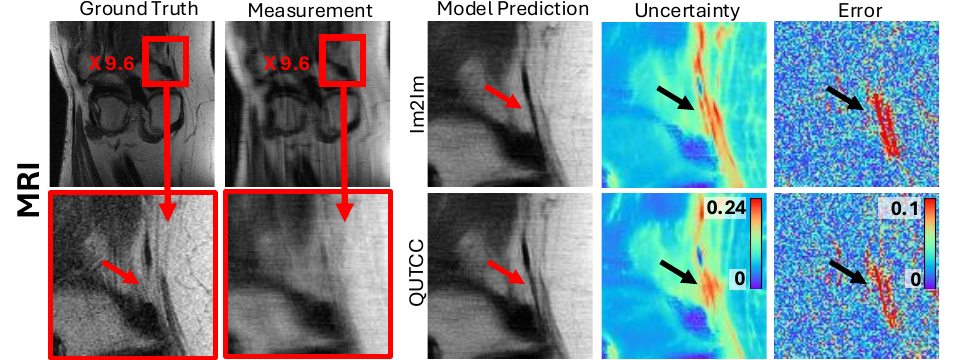}
  \caption{\textbf{Hallucination in a high uncertainty region.} 
  The predicted image and corresponding pixel-wise uncertainty map for \ours and Im2Im for accelerated MRI. The uncertainty prediction is correlated with the true error (right). Both models hallucinate a feature that is not present in the ground truth (arrow). This hallucination is co-located with a high uncertainty region.}  
  \label{fig:hallucination_visualization}
\end{figure*}
Next, we visualize the predicted pixel-wise uncertainty for Im2Im and \ours for an undersampled MRI image (Fig.~\ref{fig:hallucination_visualization}), which we compare against the true error. For both methods, high predicted uncertainty regions correspond to regions with high reconstruction error. Both models hallucinate a feature in the image reconstruction that is not present in the ground truth, as shown by arrows. For both methods, this hallucination is co-located with a high-uncertainty region in the uncertainty prediction. \ours produces a slightly narrower uncertainty interval around this region, suggesting better spatial pinpointing of high uncertainty regions. See Appendix~\ref{sup:additional_visualizations} for additional visualizations.
\subsection{Estimating the Conditional Distribution}
\begin{figure}[h]
  \centering
  \includegraphics[width=\linewidth]{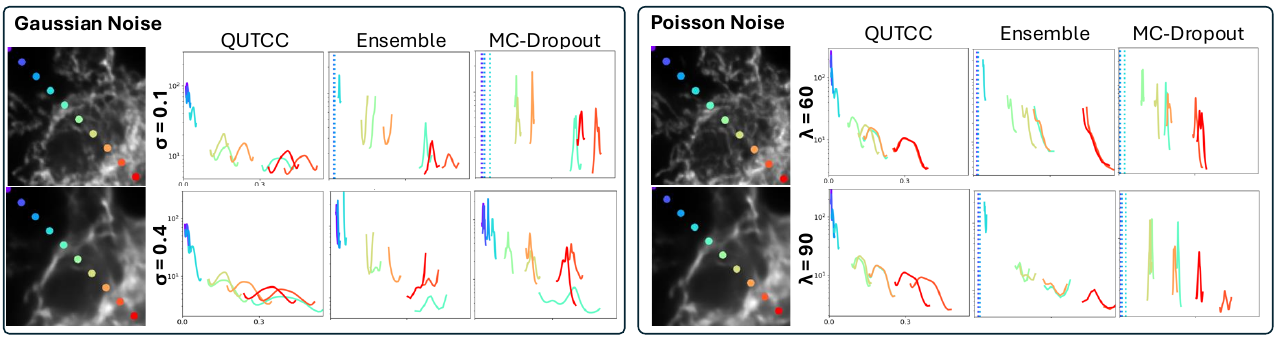}
    \caption{\textbf{\ours predicts diverse pixel-wise distributions under different noise distributions.} We compare the predicted conditional probability density functions (PDFs) of 10 representative pixels recovered from a measurement with different levels of Gaussian and Poisson noise across QUTCC, Ensemble, and MC-Dropout. Consistent with a Gaussian distribution, as $\sigma$ increases, the conditional PDF at each pixel becomes broader, reflecting the increased variance of the distribution. In contrast, in the Poisson setting, increasing $\lambda$ shifts the distribution upward, while the spread increases in proportion to the mean.}
  \label{fig:different_pdfs}
\end{figure}
Finally, we demonstrate \ours 's ability to predict a conformalized pixel-wise conditional PDF, first on a simple toy-distribution, and then on microscopy images. 

\begin{wraptable}{r}{0.48\textwidth}
%\vspace{-5pt}
\centering
\small
\setlength{\tabcolsep}{3pt}
\caption{Conditional distribution metrics.}
\label{tab:KL_divergence}
\begin{tabular}{lcccc}
\toprule
 & KL Divergence $\downarrow$ & Wasserstein $\downarrow$ \\
\midrule
MC-Dropout & $6.928 \pm 3.077$ & $0.050 \pm 0.040$  \\
Deep Ensemble & $0.307 \pm 0.046$ & $0.054 \pm 0.035$  \\
QUTCC  & $\mathbf{0.293 \pm 0.240}$ & $\mathbf{0.048 \pm 0.042}$  \\
\bottomrule
%\vspace{5pt}
\end{tabular}
\end{wraptable}
\noindent First, we consider a toy problem in which the posterior is available in closed form, as described in ShuffleFlow~\citep{li2026shuffleflow}. 
We simulate a measurement of an MNIST `3' digit through a compressive sensing matrix with additive Gaussian noise. We fit a per-pixel Gaussian prior from all the `3' digits in the MNIST dataset. Together, these give us the exact posterior marginal PDF. 
%the Following the toy example 
%\citep{li2026shuffleflow}, we consider a linear-Gaussian inverse problem in which the posterior is available in closed form: an MNIST digit is passed through a dense random linear operator with additive Gaussian noise, a per-pixel diagonal Gaussian prior is fit from real MNIST digits, and Bayesian linear-Gaussian conjugacy then gives the exact posterior mean and per-pixel variance, i.e. the true marginal PDF at every pixel. 
We compare our method against MC-Dropout (dropout kept active at inference, $S=50$ samples) and a 100-member Deep Ensemble trained with an identical, lighter-weight U-Net backbone on this toy problem. Table~\ref{tab:KL_divergence} shows the KL divergence and Wasserstein distance between each method's estimated conditional PDF and the true Gaussian marginal.
%The mean and standard deviation of the KL divergence for QUTCC, MC-Dropout, and Deep Ensemble are
%$0.293 \pm 0.240$, $6.928 \pm 3.077$, and $0.307 \pm 0.046$, respectively. The mean and standard deviation of the Wasserstein distance for QUTCC, MC-Dropout, and Deep Ensemble are $0.048 \pm 0.042$, $0.050 \pm 0.040$, and $0.054 \pm 0.035$, respectively. 
QUTCC's predicted conditional distribution is closest to the true distribution across the methods. Moreover, it does so at a lower training time than Deep Ensemble, since it requires training only a single model. For visualizations, please see Fig.~\ref{sup:toy_example} in the supplement. 
% , unlike Deep Ensembles (which require training multiple independent models) and doesn’t require multiple forward passes at different rates of dropout (unlike MC-Dropout). 

\noindent Next, we simulate measurements with varying amounts of known Gaussian and Poisson noise to test how the predicted conditional PDF varies with measurement noise. We compare our predicted conditional PDF to those predicted via Deep Ensembles and MC-Dropout. A key advantage of \ours is that it constructs these distributions by querying an arbitrary number of quantiles, enabling a flexible estimate of the predictive density. In contrast, ensemble-based and MC-Dropout methods are limited by the number of ensemble members or stochastic forward passes, which constrains the resolution of the resulting distributions. As a result, certain pixel distributions appear as dotted lines (Fig.~\ref{fig:different_pdfs}), indicating point predictions due to insufficient variability in the sampled outputs. Furthermore, these approaches typically rely on implicit Bayesian assumptions about the underlying predictive distribution (e.g., Gaussian likelihoods or approximate posterior sampling), whereas \ours is distribution-free.

In \ours, the recovered PDFs reflect the underlying noise characteristics. Under Poisson noise, the predictive distributions exhibit a pronounced skew consistent with Poisson behavior, whereas under Gaussian noise they remain approximately symmetric at higher intensity values, with asymmetry introduced near lower intensities due to clipping effects. Additionally, the predicted PDFs broaden as the noise standard deviation increases, indicating that \ours appropriately captures changes in noise magnitude. Joint reconstruction and PDF estimation could be useful to assess the underlying noise distribution present in an image, and estimating the PDF has been useful for a number of tasks in microscopy~\cite{krull2020probabilistic}.
See Appendix~\ref{sup:additional_pdf} for additional results.

\section{Limitations and Conclusion}
\label{sec:conclusion}
\ours improves existing conformal calibration methods for high-dimensional image-to-image regression problems by leveraging a learned, non-linear, and spatially-adaptive conformal calibration procedure. \ours is most beneficial over existing scalar and partition-based conformal calibration methods when the per-pixel uncertainty varies across the image, such as for signal-dependent Poisson noise. Compared to K-RCPS, which performs conformal calibration over a small, discrete number of heuristic partitions for an image (Fig~\ref{sup:diff_lambdas}), \ours produces the equivalent of continuous, adaptive partitions which are a function of the local image content.  This is made possible through simultaneous quantile regression with a quantile-embedded U-Net, enabling the conformal calibration procedure to be a function of a neural network that has learned to estimate not just a mean image, but any conditional quantile of the reconstruction. Our method achieves smaller uncertainty intervals on five imaging inverse problems than prior conformal methods and competitive performance with MC-Dropout and Deep Ensembles at a fraction of the time and compute cost, all while maintaining the same statistical coverage.  

In addition, we demonstrated a method to estimate a conformalized conditional PDF; something previous conformal methods for image-to-image regression could not do. Finally, we show several exciting examples in which regions of high predicted uncertainty are co-located with regions of high reconstruction error and structural hallucinations. We hope this line of work will help practitioners identify harmful incorrect predictions from machine learning models for scientific and medical imaging applications. 

%\ours is most beneficial over scalar (Im2Im) and partition-based (K-RCPS) conformal calibration when the per-pixel uncertainty varies across space, such as with signal-dependent Poisson noise.  \ours's learned, per-pixel scaling is designed to exploit this variation. 

Some limitations are that \ours requires paired data for both training and calibration, and offers only pixel-wise marginal coverage guarantees. Recent work attempts to relax the paired-data requirement for conformal prediction, offering a promising path forward in data-limited regimes~\cite{everink2025self, amougou2025self}. The marginal coverage guarantee is not unique to \ours but reflects a broader gap in conformal image-regression methods, which typically lack joint or spatially structured guarantees. Risks include the model incorrectly predicting low uncertainty in areas with potential model mistakes, such as for rare events or out of distribution data. However, as machine learning models improve, we hope that advances in uncertainty quantification can mitigate risks to enable more trustworthy imaging. 
% Interesting future work includes considering the effects of sample movement and distribution shifts, as well as uncertainty across multiple measurements instead of considering uncertainty for a single-frame independently. 
% Additionally, exploring the integration of quantile conditioning into multi-hypothesis prediction architectures may enable the model to capture multiple plausible outcomes while leveraging the advantages provided by the learned quantiles.
\newpage
% \section{Ethics and Reproducibility}
%Neither count toward the page limit
\textbf{Ethical Considerations:} Using machine learning methods for scientific and medical applications has inherent risk - producing realistic artifacts that are not truly present in the image can be catastrophic for discovery and medical diagnostics. As research in uncertainty quantification matures, we hope that some of these risks can be mitigated to enable more trustworthy imaging. We acknowledge and fully comply with the NeurIPS Code of Ethics.

\textbf{Reproducibility:} We describe the different imaging inverse tasks in more detail in ~\ref{sup:experiment_details}. In this section, we also include the model epochs that we used to evaluate in this paper. Additional explanation of \ours's model architecture can be found in ~\ref{sup:model_arch}, where we describe the quantile embedding in more detail. Code to train our model and reproduce the experiments shown here can be found at the code repo. 

\section{Acknowledgement}
\label{sec:Acknowledgement}
The authors would like to thank Anastasios Angelopoulous for reviewing and providing feedback on our manuscript. The authors would like to thank Nora Gera, Benjamin Li, Gabriel Morete, and Adrian Hayler for their help running preliminary experiments and for valuable feedback on the statistical analysis. Research reported in this publication was supported by an Amazon Research Award, Fall 2024. We also thank the NeurIPS 2026 Financial Assistance Program for travel assistance. 

{\small
\bibliographystyle{unsrt}
\bibliography{example_paper}

@inproceedings{angelopoulos2022image,
  title={Image-to-image regression with distribution-free uncertainty quantification and applications in imaging},
  author={Angelopoulos, Anastasios N and Kohli, Amit Pal and Bates, Stephen and Jordan, Michael and Malik, Jitendra and Alshaabi, Thayer and Upadhyayula, Srigokul and Romano, Yaniv},
  booktitle={International Conference on Machine Learning},
  pages={717--730},
  year={2022},
  organization={PMLR}
}

@misc{ongie2020deeplearningtechniquesinverse,
      title={Deep Learning Techniques for Inverse Problems in Imaging}, 
      author={Gregory Ongie and Ajil Jalal and Christopher A. Metzler and Richard G. Baraniuk and Alexandros G. Dimakis and Rebecca Willett},
      year={2020},
      eprint={2005.06001},
      archivePrefix={arXiv},
      primaryClass={eess.IV},
      url={https://arxiv.org/abs/2005.06001}, 
}

@article{alshardan2024deep,
  title={Deep learning solutions for inverse problems in advanced biomedical image analysis on disease detection},
  author={Alshardan, Amal and Mahgoub, Hany and Alruwais, Nuha and Darem, Abdulbasit A and Almukadi, Wafa Sulaiman and Mohamed, Abdullah},
  journal={Scientific Reports},
  volume={14},
  number={1},
  pages={18478},
  year={2024},
  publisher={Nature Publishing Group UK London}
}

@article{gupta2023topology,
  title={Topology-aware uncertainty for image segmentation},
  author={Gupta, Saumya and Zhang, Yikai and Hu, Xiaoling and Prasanna, Prateek and Chen, Chao},
  journal={Advances in Neural Information Processing Systems},
  volume={36},
  pages={8186--8207},
  year={2023}
}

@article{romano2019conformalized,
  title={Conformalized quantile regression},
  author={Romano, Yaniv and Patterson, Evan and Candes, Emmanuel},
  journal={Advances in neural information processing systems},
  volume={32},
  year={2019}
}

@article{chung2021beyond,
  title={Beyond pinball loss: Quantile methods for calibrated uncertainty quantification},
  author={Chung, Youngseog and Neiswanger, Willie and Char, Ian and Schneider, Jeff},
  journal={Advances in Neural Information Processing Systems},
  volume={34},
  pages={10971--10984},
  year={2021}
}

@article{ye2025learned,
  title={Learned, uncertainty-driven adaptive acquisition for photon-efficient scanning microscopy},
  author={Ye, Cassandra Tong and Han, Jiashu and Liu, Kunzan and Angelopoulos, Anastasios and Griffith, Linda and Monakhova, Kristina and You, Sixian},
  journal={Optics Express},
  volume={33},
  number={6},
  pages={12269--12287},
  year={2025},
  publisher={Optica Publishing Group}
}

@inproceedings{gal2016dropout,
  title={Dropout as a bayesian approximation: Representing model uncertainty in deep learning},
  author={Gal, Yarin and Ghahramani, Zoubin},
  booktitle={international conference on machine learning},
  pages={1050--1059},
  year={2016},
  organization={PMLR}
}

@article{lakshminarayanan2017simple,
  title={Simple and scalable predictive uncertainty estimation using deep ensembles},
  author={Lakshminarayanan, Balaji and Pritzel, Alexander and Blundell, Charles},
  journal={Advances in neural information processing systems},
  volume={30},
  year={2017}
}

@article{zhang2023unified,
  title={A unified conditional framework for diffusion-based image restoration},
  author={Zhang, Yi and Shi, Xiaoyu and Li, Dasong and Wang, Xiaogang and Wang, Jian and Li, Hongsheng},
  journal={Advances in Neural Information Processing Systems},
  volume={36},
  pages={49703--49714},
  year={2023}
}

@article{angelopoulos2021gentle,
  title={A gentle introduction to conformal prediction and distribution-free uncertainty quantification},
  author={Angelopoulos, Anastasios N and Bates, Stephen},
  journal={arXiv preprint arXiv:2107.07511},
  year={2021}
}

@article{correia2024information,
  title={An Information Theoretic Perspective on Conformal Prediction},
  author={Correia, Alvaro HC and Massoli, Fabio Valerio and Louizos, Christos and Behboodi, Arash},
  journal={arXiv preprint arXiv:2405.02140},
  year={2024}
}

@inproceedings{wen2024task,
  title={Task-Driven Uncertainty Quantification in Inverse Problems via Conformal Prediction},
  author={Wen, Jeffrey and Ahmad, Rizwan and Schniter, Philip},
  booktitle={European Conference on Computer Vision},
  pages={182--199},
  year={2024},
  organization={Springer}
}

@article{rodrigues2020beyond,
  title={Beyond expectation: Deep joint mean and quantile regression for spatiotemporal problems},
  author={Rodrigues, Filipe and Pereira, Francisco C},
  journal={IEEE transactions on neural networks and learning systems},
  volume={31},
  number={12},
  pages={5377--5389},
  year={2020},
  publisher={IEEE}
}

@article{nehme2023uncertainty,
  title={Uncertainty quantification via neural posterior principal components},
  author={Nehme, Elias and Yair, Omer and Michaeli, Tomer},
  journal={Advances in Neural Information Processing Systems},
  volume={36},
  pages={37128--37141},
  year={2023}
}

@article{belhasin2023principal,
  title={Principal uncertainty quantification with spatial correlation for image restoration problems},
  author={Belhasin, Omer and Romano, Yaniv and Freedman, Daniel and Rivlin, Ehud and Elad, Michael},
  journal={IEEE Transactions on Pattern Analysis and Machine Intelligence},
  volume={46},
  number={5},
  pages={3321--3333},
  year={2023},
  publisher={IEEE}
}

@inproceedings{yair2024uncertainty,
  title={Uncertainty visualization via low-dimensional posterior projections},
  author={Yair, Omer and Nehme, Elias and Michaeli, Tomer},
  booktitle={Proceedings of the IEEE/CVF Conference on Computer Vision and Pattern Recognition},
  pages={11041--11051},
  year={2024}
}

@article{xie2024boosted,
  title={Boosted conformal prediction intervals},
  author={Xie, Ran and Barber, Rina and Candes, Emmanuel},
  journal={Advances in Neural Information Processing Systems},
  volume={37},
  pages={71868--71899},
  year={2024}
}

@article{koenker1978regression,
  title={Regression quantiles},
  author={Koenker, Roger and Bassett Jr, Gilbert},
  journal={Econometrica: journal of the Econometric Society},
  pages={33--50},
  year={1978},
  publisher={JSTOR}
}

@article{barbastathis2019use,
  title={On the use of deep learning for computational imaging},
  author={Barbastathis, George and Ozcan, Aydogan and Situ, Guohai},
  journal={Optica},
  volume={6},
  number={8},
  pages={921--943},
  year={2019},
  publisher={Optical Society of America}
}

@article{angelopoulos2020uncertainty,
  title={Uncertainty sets for image classifiers using conformal prediction},
  author={Angelopoulos, Anastasios and Bates, Stephen and Malik, Jitendra and Jordan, Michael I},
  journal={arXiv preprint arXiv:2009.14193},
  year={2020}
}

@article{quach2023conformal,
  title={Conformal language modeling},
  author={Quach, Victor and Fisch, Adam and Schuster, Tal and Yala, Adam and Sohn, Jae Ho and Jaakkola, Tommi S and Barzilay, Regina},
  journal={arXiv preprint arXiv:2306.10193},
  year={2023}
}

@article{ding2023class,
  title={Class-conditional conformal prediction with many classes},
  author={Ding, Tiffany and Angelopoulos, Anastasios and Bates, Stephen and Jordan, Michael and Tibshirani, Ryan J},
  journal={Advances in neural information processing systems},
  volume={36},
  pages={64555--64576},
  year={2023}
}

@article{lindemann2023safe,
  title={Safe planning in dynamic environments using conformal prediction},
  author={Lindemann, Lars and Cleaveland, Matthew and Shim, Gihyun and Pappas, George J},
  journal={IEEE Robotics and Automation Letters},
  volume={8},
  number={8},
  pages={5116--5123},
  year={2023},
  publisher={IEEE}
}

@article{fannjiang2022conformal,
  title={Conformal prediction under feedback covariate shift for biomolecular design},
  author={Fannjiang, Clara and Bates, Stephen and Angelopoulos, Anastasios N and Listgarten, Jennifer and Jordan, Michael I},
  journal={Proceedings of the National Academy of Sciences},
  volume={119},
  number={43},
  pages={e2204569119},
  year={2022},
  publisher={National Academy of Sciences}
}

@inproceedings{lekeufack2024conformal,
  title={Conformal decision theory: Safe autonomous decisions from imperfect predictions},
  author={Lekeufack, Jordan and Angelopoulos, Anastasios N and Bajcsy, Andrea and Jordan, Michael I and Malik, Jitendra},
  booktitle={2024 IEEE International Conference on Robotics and Automation (ICRA)},
  pages={11668--11675},
  year={2024},
  organization={IEEE}
}

@article{campos2024conformal,
  title={Conformal prediction for natural language processing: A survey},
  author={Campos, Margarida and Farinhas, Ant{\'o}nio and Zerva, Chrysoula and Figueiredo, M{\'a}rio AT and Martins, Andr{\'e} FT},
  journal={Transactions of the Association for Computational Linguistics},
  volume={12},
  pages={1497--1516},
  year={2024},
  publisher={MIT Press 255 Main Street, 9th Floor, Cambridge, Massachusetts 02142, USA~…}
}

@article{koenker2001quantile,
  title={Quantile regression},
  author={Koenker, Roger and Hallock, Kevin F},
  journal={Journal of economic perspectives},
  volume={15},
  number={4},
  pages={143--156},
  year={2001},
  publisher={American Economic Association}
}

@article{steinwart2011estimating,
  title={Estimating conditional quantiles with the help of the pinball loss},
  author={Steinwart, Ingo and Christmann, Andreas},
  year={2011}
}

@inproceedings{dabney2018implicit,
  title={Implicit quantile networks for distributional reinforcement learning},
  author={Dabney, Will and Ostrovski, Georg and Silver, David and Munos, R{\'e}mi},
  booktitle={International conference on machine learning},
  pages={1096--1105},
  year={2018},
  organization={PMLR}
}

@article{oktay2018attention,
  title={Attention u-net: Learning where to look for the pancreas},
  author={Oktay, Ozan and Schlemper, Jo and Folgoc, Loic Le and Lee, Matthew and Heinrich, Mattias and Misawa, Kazunari and Mori, Kensaku and McDonagh, Steven and Hammerla, Nils Y and Kainz, Bernhard and others},
  journal={arXiv preprint arXiv:1804.03999},
  year={2018}
}

@inproceedings{ronneberger2015u,
  title={U-net: Convolutional networks for biomedical image segmentation},
  author={Ronneberger, Olaf and Fischer, Philipp and Brox, Thomas},
  booktitle={Medical image computing and computer-assisted intervention--MICCAI 2015: 18th international conference, Munich, Germany, October 5-9, 2015, proceedings, part III 18},
  pages={234--241},
  year={2015},
  organization={Springer}
}

@article{ho2020denoising,
  title={Denoising diffusion probabilistic models},
  author={Ho, Jonathan and Jain, Ajay and Abbeel, Pieter},
  journal={Advances in neural information processing systems},
  volume={33},
  pages={6840--6851},
  year={2020}
}

@inproceedings{vovk2012conditional,
  title={Conditional validity of inductive conformal predictors},
  author={Vovk, Vladimir},
  booktitle={Asian conference on machine learning},
  pages={475--490},
  year={2012},
  organization={PMLR}
}

@article{zbontar2018fastmri,
  title={fastMRI: An open dataset and benchmarks for accelerated MRI},
  author={Zbontar, Jure and Knoll, Florian and Sriram, Anuroop and Murrell, Tullie and Huang, Zhengnan and Muckley, Matthew J and Defazio, Aaron and Stern, Ruben and Johnson, Patricia and Bruno, Mary and others},
  journal={arXiv preprint arXiv:1811.08839},
  year={2018}
}

@article{pinkard2024berkeley,
  title={The berkeley single cell computational microscopy (bsccm) dataset},
  author={Pinkard, Henry and Liu, Cherry and Nyatigo, Fanice and Fletcher, Daniel A and Waller, Laura},
  journal={arXiv preprint arXiv:2402.06191},
  year={2024}
}

@inproceedings{zhang2019poisson,
  title={A poisson-gaussian denoising dataset with real fluorescence microscopy images},
  author={Zhang, Yide and Zhu, Yinhao and Nichols, Evan and Wang, Qingfei and Zhang, Siyuan and Smith, Cody and Howard, Scott},
  booktitle={Proceedings of the IEEE/CVF Conference on Computer Vision and Pattern Recognition},
  pages={11710--11718},
  year={2019}
}

@article{knoll2020fastmri,
  title={fastMRI: A publicly available raw k-space and DICOM dataset of knee images for accelerated MR image reconstruction using machine learning},
  author={Knoll, Florian and Zbontar, Jure and Sriram, Anuroop and Muckley, Matthew J and Bruno, Mary and Defazio, Aaron and Parente, Marc and Geras, Krzysztof J and Katsnelson, Joe and Chandarana, Hersh and others},
  journal={Radiology: Artificial Intelligence},
  volume={2},
  number={1},
  pages={e190007},
  year={2020},
  publisher={Radiological Society of North America}
}

@article{kingma2014adam,
  title={Adam: A method for stochastic optimization},
  author={Kingma, Diederik P and Ba, Jimmy},
  journal={arXiv preprint arXiv:1412.6980},
  year={2014}
}

@inproceedings{rupprecht2017learning,
  title={Learning in an uncertain world: Representing ambiguity through multiple hypotheses},
  author={Rupprecht, Christian and Laina, Iro and DiPietro, Robert and Baust, Maximilian and Tombari, Federico and Navab, Nassir and Hager, Gregory D},
  booktitle={Proceedings of the IEEE international conference on computer vision},
  pages={3591--3600},
  year={2017}
}

@inproceedings{ilg2018uncertainty,
  title={Uncertainty estimates and multi-hypotheses networks for optical flow},
  author={Ilg, Eddy and Cicek, Ozgun and Galesso, Silvio and Klein, Aaron and Makansi, Osama and Hutter, Frank and Brox, Thomas},
  booktitle={Proceedings of the European Conference on Computer Vision (ECCV)},
  pages={652--667},
  year={2018}
}

@inproceedings{everink2025self,
  title={Self-supervised conformal prediction for uncertainty quantification in imaging problems},
  author={Everink, Jasper Marijn and Amougou, Bernardin Tamo and Pereyra, Marcelo},
  booktitle={International Conference on Scale Space and Variational Methods in Computer Vision},
  pages={108--118},
  year={2025},
  organization={Springer}
}

@inproceedings{amougou2025self,
  title={Self-supervised conformal prediction for uncertainty quantification in Poisson imaging problems},
  author={Amougou, Bernardin Tamo and Pereyra, Marcelo and Pascal, Barbara},
  booktitle={2025 IEEE Statistical Signal Processing Workshop (SSP)},
  pages={76--80},
  year={2025},
  organization={IEEE}
}

@article{nehme2024hierarchical,
  title={Hierarchical uncertainty exploration via feedforward posterior trees},
  author={Nehme, Elias and Mulayoff, Rotem and Michaeli, Tomer},
  journal={Advances in Neural Information Processing Systems},
  volume={37},
  pages={125142--125191},
  year={2024}
}

@article{krull2020probabilistic,
  title={Probabilistic noise2void: Unsupervised content-aware denoising},
  author={Krull, Alexander and Vi{\v{c}}ar, Tom{\'a}{\v{s}} and Prakash, Mangal and Lalit, Manan and Jug, Florian},
  journal={Frontiers in Computer Science},
  volume={2},
  pages={5},
  year={2020},
  publisher={Frontiers Media SA}
}

@inproceedings{teneggi2023trust,
  title={How to trust your diffusion model: A convex optimization approach to conformal risk control},
  author={Teneggi, Jacopo and Tivnan, Matthew and Stayman, Web and Sulam, Jeremias},
  booktitle={International Conference on Machine Learning},
  pages={33940--33960},
  year={2023},
  organization={PMLR}
}

@inproceedings{ostrovski2018autoregressive,
  title={Autoregressive quantile networks for generative modeling},
  author={Ostrovski, Georg and Dabney, Will and Munos, R{\'e}mi},
  booktitle={International Conference on Machine Learning},
  pages={3936--3945},
  year={2018},
  organization={PMLR}
}

@inproceedings{vovk2017nonparametric,
  title={Nonparametric predictive distributions based on conformal prediction},
  author={Vovk, Vladimir and Shen, Jieli and Manokhin, Valery and Xie, Min-ge},
  booktitle={Conformal and probabilistic prediction and applications},
  pages={82--102},
  year={2017},
  organization={PMLR}
}

@inproceedings{vovk2018conformal,
  title={Conformal predictive distributions with kernels},
  author={Vovk, Vladimir and Nouretdinov, Ilia and Manokhin, Valery and Gammerman, Alex},
  booktitle={Braverman Readings in Machine Learning. Key Ideas from Inception to Current State: International Conference Commemorating the 40th Anniversary of Emmanuil Braverman's Decease, Boston, MA, USA, April 28-30, 2017, Invited Talks},
  pages={103--121},
  year={2018},
  organization={Springer}
}

@article{manor2023posterior,
  title={On the posterior distribution in denoising: Application to uncertainty quantification},
  author={Manor, Hila and Michaeli, Tomer},
  journal={arXiv preprint arXiv:2309.13598},
  year={2023}
}

@article{blei2017variational,
  title={Variational inference: A review for statisticians},
  author={Blei, David M and Kucukelbir, Alp and McAuliffe, Jon D},
  journal={Journal of the American statistical Association},
  volume={112},
  number={518},
  pages={859--877},
  year={2017},
  publisher={Taylor \& Francis}
}

@book{brooks2011handbook,
  title={Handbook of markov chain monte carlo},
  author={Brooks, Steve and Gelman, Andrew and Jones, Galin and Meng, Xiao-Li},
  year={2011},
  publisher={CRC press}
}

@article{gal2015bayesian,
  title={Bayesian convolutional neural networks with Bernoulli approximate variational inference},
  author={Gal, Yarin and Ghahramani, Zoubin},
  journal={arXiv preprint arXiv:1506.02158},
  year={2015}
}

@inproceedings{feng2023score,
  title={Score-based diffusion models as principled priors for inverse imaging},
  author={Feng, Berthy T and Smith, Jamie and Rubinstein, Michael and Chang, Huiwen and Bouman, Katherine L and Freeman, William T},
  booktitle={Proceedings of the IEEE/CVF International Conference on Computer Vision},
  pages={10520--10531},
  year={2023}
}

@inproceedings{dabney2018distributional,
  title={Distributional reinforcement learning with quantile regression},
  author={Dabney, Will and Rowland, Mark and Bellemare, Marc and Munos, R{\'e}mi},
  booktitle={Proceedings of the AAAI conference on artificial intelligence},
  volume={32},
  number={1},
  year={2018}
}

@article{weigert2018content,
  title={Content-aware image restoration: pushing the limits of fluorescence microscopy},
  author={Weigert, Martin and Schmidt, Uwe and Boothe, Tobias and M{\"u}ller, Andreas and Dibrov, Alexandr and Jain, Akanksha and Wilhelm, Benjamin and Schmidt, Deborah and Broaddus, Coleman and Culley, Si{\^a}n and others},
  journal={Nature methods},
  volume={15},
  number={12},
  pages={1090--1097},
  year={2018},
  publisher={Nature Publishing Group US New York}
}

@article{bates2021distribution,
  title={Distribution-free, risk-controlling prediction sets},
  author={Bates, Stephen and Angelopoulos, Anastasios and Lei, Lihua and Malik, Jitendra and Jordan, Michael},
  journal={Journal of the ACM (JACM)},
  volume={68},
  number={6},
  pages={1--34},
  year={2021},
  publisher={ACM New York, NY}
}

@inproceedings{li2026shuffleflow,
  title={ShuffleFlow: Scalable Posterior Inference for Bayesian Inverse Imaging},
  author={Li, Tianao and Starkenburg, Tjitske and Sun, Yu and Alexander, Emma},
  booktitle={2026 IEEE International Conference on Computational Photography (ICCP)},
  pages={1--11},
  year={2026},
  organization={IEEE}
}

@article{otsu1979threshold,
  title={A threshold selection method from gray-level histograms},
  author={Otsu, Nobuyuki},
  journal={IEEE transactions on systems, man, and cybernetics},
  volume={9},
  number={1},
  pages={62--66},
  year={1979},
  publisher={IEEE}
}

@inproceedings{teneggi2025conformal,
  title={Conformal risk control for semantic uncertainty quantification in computed tomography},
  author={Teneggi, Jacopo and Stayman, J Webster and Sulam, Jeremias},
  booktitle={International Conference on Medical Image Computing and Computer-Assisted Intervention},
  pages={45--55},
  year={2025},
  organization={Springer}
}
}
%%%%%%%%%%%%%%%%%%%%%%%%%%%%%%%%%%%%%%%%%%%%%%%%%%%%%%%%%%%%

\appendix
\newpage
\section{Appendix}
These supplementary materials contain additional information to improve the reproducibility of the paper, as well as several additional analyses and experimental results. Code for training and analysis of QUTCC can be found at our repository. Section~\ref{related_work} details related work to our method. Section~\ref{sup:model_arch} details the network architecture used for \ours. Section~\ref{sup:model_analysis} provides additional comparisons against the original formulation of Im2Im-UQ without our network modifications and analysis on quantile crossing. Section~\ref{sup:model_analysis} also includes analysis on the size-stratified risk, more visualizations of our uncertainty predictions across the different inverse problems, and a visualization of how \ours produces narrower uncertainty intervals through asymmetry. Section~\ref{sec:adaptive_scaling_vs_krcps} provides a comparison of the scaling used in Im2Im, K-RCPS, and \ours. Sec.~\ref{sup:additional_pdf} provides several visualizations of \ours's conditional distribution predictions, including an example showing how we conformalize the quantiles and the PDF prediction as a function of measurement noise. Finally, Section~\ref{sup:experiment_details} provides more details about the different inverse problems used throughout the paper and their associated datasets.  

\subsection{Related Work}
\label{related_work}
\subsubsection{Quantile Regression}
Quantile regression is a general approach to estimate the conditional quantiles of a target distribution ~\citep{koenker1978regression, koenker2001quantile}. This is often accomplished by leveraging an asymmetric loss function, called pinball loss (Fig.~\ref{fig:architecture}a , Eq.~\ref{eq:pinball_loss}), tailored to the specified quantile level~\citep{steinwart2011estimating}. The estimated intervals obtained by quantile regression do not have formal guarantees on their own, but can be paired with conformal prediction to obtain coverage guarantees~\citep{romano2019conformalized}. Learning quantiles during neural network training can improve predictive performance through regularization while enabling uncertainty estimation~\citep{rodrigues2020beyond}. 
% One limitation of quantile-based methods is `quantile crossing,' where independently trained quantiles violate their natural ordering, with lower quantiles exceeding higher ones \citep{das2019quantile}. Simultaneous quantile prediction, training a single network to predict multiple quantiles, can mitigate this issue while also enabling estimation of the entire conditional distribution~\citep{sangnier2016joint, liu2011simultaneous, tagasovska2019single, rodrigues2020beyond}. 
In our work, we leverage a single-network with shared parameters for simultaneous quantile prediction. We embed the quantile level as an explicit input parameter into a U-Net, which is well-suited for a variety of image regression and imaging inverse tasks. Embedding a notion of quantiles into deep learning architectures has been explored in reinforcement learning and generative modeling \cite{dabney2018implicit, ostrovski2018autoregressive}, but its application to predicting uncertainty bounds for image regression tasks remains largely unexplored. We are the first to demonstrate that a single network trained for simultaneous quantile prediction can predict conformally calibrated uncertainty intervals, while mitigating quantile crossing, for imaging inverse problems.
% Furthermore, we pair our network with conformal prediction to achieve coverage guarantees. We show that our network mitigates the issue of quantile crossing while also maintaining overall prediction accuracy. We are the first to demonstrate that a single network trained for simultaneous quantile prediction can predict conformally calibrated uncertainty intervals for imaging inverse problems.
\subsubsection{Predicting Tighter or More Interpretable Bounds}
Achieving smaller interval lengths without sacrificing coverage guarantees reflects increased confidence in the model’s predictions. Producing smaller intervals is a common objective across uncertainty estimation methods, not just conformal prediction~\citep{xie2024boosted}. Several approaches have been proposed to enhance conformal prediction by targeting user-specified properties such as reduced interval length or improved conditional coverage~\citep{xie2024boosted,chung2021beyond}; however, to date, none of these techniques have been applied to imaging tasks. On the other hand, several methods aim to improve the interpretability of uncertainty prediction for imaging tasks by moving away from per-pixel uncertainty estimates. These methods leverage principal components, posterior projected distribution, and spatial/topological relationships \citep{nehme2023uncertainty, yair2024uncertainty, belhasin2023principal, gupta2023topology} to predict uncertainty in a more interpretable way. A few exciting directions include using conformal prediction for semantic uncertainty quantification in applications such as computed tomography~\cite{teneggi2025conformal}, and task-driven uncertainty quantification~\cite{wen2024task}, but this relies on labeled data which is not always available for imaging problems. In addition, studies in multi-hypothesis uncertainty estimation have investigated multi-head and mixture-based networks that predict multiple candidate hypotheses to explore the space of potential solutions instead of presenting a single candidate reconstruction \citep{rupprecht2017learning, ilg2018uncertainty, nehme2024hierarchical}. However, without incorporating conformal prediction, these methods lack statistical guarantees. 
% Several methods pair inverse problems with downstream tasks, such as classification, to estimate the uncertainty in a more interpretable way~\citep{cheung2024metric, wen2024task}, and others represent uncertainty in a semantically-meaningful latent space~\citep{sankaranarayanan2022semantic}. While these methods are promising, they are less general and often tied to a specific application. Additional work have also investigated risk controlling prediction sets for image regression tasks in generative and medical tasks \citep{fischer2024subgroup, teneggi2023trust}.
We present a more general method that can predict uncertainty for any imaging inverse problem, without task-specific labels, while achieving smaller uncertainty interval lengths than previous image-to-image regression methods. 
\subsubsection{Predicting the Conditional Distribution}
In some scenarios, such as probabilistic image denoising~\cite{krull2020probabilistic, manor2023posterior}, predicting a full conditional probability distribution, $p(\mathbf x| \mathbf y)$, could have value over predicting a confidence interval.  This can be accomplished with classic Bayesian approaches like variational inference~\cite{blei2017variational, gal2015bayesian, feng2023score} and Markov chain Monte Carlo~\cite{brooks2011handbook}, but these either require strong distributional assumptions or are impractically slow. There has been some theoretical work in extending conformal prediction methods to construct full nonparametric predictive distributions with guaranteed coverage properties~\citep{vovk2017nonparametric, vovk2018conformal}. In addition, conditional distributions have been estimated using a quantile loss for problems in distributional reinforcement learning for Atari games~\citep{dabney2018distributional,dabney2018implicit}. Our work extends the ideas from conformal predictive distributions to multi-dimensional imaging data and provides a practical way to apply these ideas to image regression and imaging inverse problems for the first time. As opposed to Bayesian methods, this approach is distribution-free, quick, and maintains conformal guarantees for the predicted conditional distribution.

\subsection{QUTCC Model Architecture}
\label{sup:model_arch}
Our method, \ours, is based on a U-Net backbone augmented with self-attention mechanisms, where quantile embeddings are propagated through the self-attention layers to guide the network's quantile predictions. This is inspired by the architecture of U-Nets used for diffusion models, which include a time-embedding~\citep{ho2020denoising,zhang2023unified}. In our design, the target quantile level is embedded as a continuous scalar, analogous to the time-step embeddings in diffusion models. A core part of this architecture is the integration of self-attention layers within the U-Net, implemented as \texttt{AttentionBlock} modules. These blocks allow the model to capture global dependencies across spatial dimensions, enabling each pixel or feature location to attend to all other locations within its feature map. Specifically, attention layers are incorporated at various downsampling resolutions within both the encoder and decoder paths, as well as at the bottleneck of the U-Net.

The ability to condition the model's output on a given quantile is achieved through the quantile embedding mechanism. A quantile value between (0, 1) is chosen for each image sample at each iteration and then transformed into a high-dimensional vector representation using a positional encoding scheme, employing sinusoidal functions to create generalizable embeddings. This initial embedding is then processed by a small multi-layer perceptron, which is then used throughout to condition the network. By introducing this conditioning, the model learns to generate outputs that are responsive to the entire range of quantile levels. This mechanism works in tandem with the self-attention layers, which are specified using the \texttt{attention\_resolutions} parameter.

In our experiments, \ours was trained using \texttt{attention\_resolutions} configured as $[16, 8, 4, 2, 1]$. This means for 512 x 512 images, using the specified configuration results in attention layers within the encoder path (at resolutions 512x512, 256x256, 128x128, 64x64, and 32x32), one central attention layer in the middle block (at 8x8 resolution), and additional attention layers distributed across the decoder path (at resolutions 32x32, 64x64, 128x128, 256x256, and 512x512). The \texttt{attention\_resolutions} parameter dictates the spatial scales at which these attention mechanisms are introduced. Further model configurations can be found in \texttt{models/model\_config.yaml}.

\subsubsection{Deep Ensemble Network Training}
We train a Deep Ensemble~\cite{lakshminarayanan2017simple} baseline with $M=10$ independently initialized U-Net predictors. Each member $f_{\theta_m}$ uses the same backbone as the image-to-image and QUTCC models, with a two-channel output head. Each model predicts a Gaussian distribution over the output, producing per-pixel mean $\mu_m$ and variance $\sigma_m^2$ via a two-channel output head, with positivity enforced using a softplus activation and a minimum floor of $10^{-6}$. We train each member for 50 epochs.

Each member is optimized independently with the Gaussian negative log-likelihood, using Adam with learning rate $10^{-4}$, $\beta=(0.9,0.99)$, and the same one-cycle schedule used for the other models. At inference, predictions from all $M$ members are aggregated to compute the ensemble predictive mean and variance via the law of total expectation and variance. Then, we apply the same conformal calibration protocol used in all other baselines.

\subsubsection{MC-Dropout Network Training}
We train a single U-Net with dropout layers inserted throughout the network at rate $p$, following \cite{gal2016dropout}. During training, dropout is active and the model is optimized with mean squared error using Adam with learning rate $10^{-4}$, $\beta=(0.9,0.99)$, and the same one-cycle learning-rate schedule as the other learned baselines. At inference, dropout remains active and each stochastic forward pass is treated as a sample from an approximate posterior. We draw $S=50$ Monte Carlo samples $\{\hat{y}_s(x)\}_{s=1}^S$ and estimate
\[
\hat{\mu}(x)=\frac{1}{S}\sum_{s=1}^S \hat{y}_s(x),\qquad
\hat{\sigma}^2(x)=\frac{1}{S}\sum_{s=1}^S(\hat{y}_s(x)-\hat{\mu}(x))^2 + \tau^{-1}.
\] We then conformally calibrate the resulting bounds. This procedure yields a practical approximation to Bayesian inference under a Gaussian process prior.

\subsection{Model Analysis}
\label{sup:model_analysis}
In Fig.~\ref{sec:results}, we restricted our comparisons to the Im2Im-Deep (which we simply refer to as Im2Im) and \ours models. The original Im2Im-UQ model was excluded due to its comparatively shallow architecture, resulting in decreased performance.  However, for the subsequent analysis, we reintroduce Im2Im-UQ for completeness. In this section, we assess the model's mean predictive performance and the quantile crossing occurrences. 

\subsubsection{Prediction Performance}
Does \ours produce tighter intervals because it is simply a better image prediction network? To investigate this, we compare the predictive performance of MC-Dropout, Deep Ensemble, Im2Im-UQ, Im2Im-Deep, Im2Im-Deep-Median, and \ours. Im2Im-Deep-Median is a variant of Im2Im-Deep that predicts the median rather than the mean. Since QUTCC’s estimates are centered on the median, we train this median-predicting version of Im2Im-Deep to ensure a fair comparison. In Table~\ref{tbl:Mean_model_peformance}, we compare model performance using standard image reconstruction metrics: MSE, SSIM, PSNR, and LPIPS. For \ours, the mean prediction was obtained by setting the quantile level to $q = 0.5$. The results indicate that all models achieve nearly identical performance in terms of MSE, with only minor differences observed in SSIM, PSNR, and LPIPS. These variations are not substantial enough to suggest that \ours provides a significantly better mean prediction. These findings suggest that \ours’s improved uncertainty quantification predictions are not attributed to better mean prediction performance. Rather, its ability to more effectively characterize uncertainty appears to come from the explicit learning of quantiles during training.
\sisetup{
  scientific-notation=false,
  exponent-product=\cdot,
  output-exponent-marker=\text{e},
  round-mode=figures,
  round-precision=2
}

\begin{table}[t]
\caption{Image reconstruction performance. Arrows indicate the direction of better performance.}
\centering
\papertablefont
\setlength{\tabcolsep}{2.5pt}
\begin{tabular*}{\textwidth}{@{\extracolsep{\fill}}llccccc@{}}
\toprule
\textbf{Metric} & \textbf{Model} & \textbf{MRI} & \textbf{QPI} & \textbf{Gaussian} & \textbf{Poisson} & \textbf{Real Noise} \\
\midrule

\multirow{6}{*}{MSE ($\downarrow$)}
  & MC-Drop.  & 0.002 $\pm$ 0.002 & 0.001 $\pm$ 0.001
  & 0.001 $\pm$ 0.001 & 0.0002 $\pm$ 0.0002 & 0.0002 $\pm$ 0.0001 \\
  & Deep Ens. & 0.001 $\pm$ 0.002 & 0.0002 $\pm$ 0.0001
  & 0.001 $\pm$ 0.001 & 0.0003 $\pm$ 0.0003 & 0.0001 $\pm$ 0.0001 \\
  & Im2Im-UQ    & 0.003 $\pm$ 0.002 & 0.0006 $\pm$ 0.0005 &
  0.0007 $\pm$ 0.0006 & 0.0003 $\pm$ 0.0003 & 0.0030 $\pm$ 0.0004 \\
  & Im2Im-Deep  & 0.001 $\pm$ 0.002 & 0.0004 $\pm$ 0.0003 &
  0.0006 $\pm$ 0.0006 & 0.0003 $\pm$ 0.0002 & 0.0004 $\pm$ 0.0002 \\ 
& Im2Im-Median & 0.001 $\pm$ 0.002 & 0.0003 $\pm$ 0.0002
& 0.0006 $\pm$ 0.0005 & 0.0003 $\pm$ 0.0002 & 0.0006 $\pm$ 0.0004 \\
  & QUTCC       & 0.001 $\pm$ 0.002 & 0.0004 $\pm$ 0.0003 &
  0.0006 $\pm$ 0.0005 & 0.0003 $\pm$ 0.0003 & 0.0002 $\pm$ 0.0001 \\
\addlinespace

\multirow{6}{*}{SSIM ($\uparrow$)}
  & MC-Drop.  & 0.684 $\pm$ 0.129 & 0.941 $\pm$ 0.017
  & 0.810 $\pm$ 0.124 & 0.941 $\pm$ 0.033 & 0.957 $\pm$ 0.004 \\
  & Deep Ens. & 0.717 $\pm$ 0.141 & 0.965 $\pm$ 0.008
  & 0.806 $\pm$ 0.126 & 0.933 $\pm$ 0.038 & 0.967 $\pm$ 0.006 \\
  & Im2Im-UQ    & 0.668 $\pm$ 0.127 & 0.949 $\pm$ 0.017 &
  0.852 $\pm$ 0.102 & 0.931 $\pm$ 0.038 & 0.803 $\pm$ 0.016 \\
  & Im2Im-Deep  & 0.707 $\pm$ 0.139 & 0.961 $\pm$ 0.010 &
  0.856 $\pm$ 0.107 & 0.937 $\pm$ 0.035 & 0.959 $\pm$ 0.006 \\
& Im2Im-Median & 0.708 $\pm$ 0.139 & 0.961 $\pm$ 0.010
& 0.865 $\pm$ 0.101 & 0.932 $\pm$ 0.039 & 0.952 $\pm$ 0.013 \\ 
  & QUTCC       & 0.708 $\pm$ 0.139 & 0.959 $\pm$ 0.010
  & 0.865 $\pm$ 0.102 & 0.941 $\pm$ 0.036 & 0.957 $\pm$ 0.008 \\
\addlinespace

\multirow{6}{*}{PSNR ($\uparrow$)}
  & MC-Drop.  & 28.381 $\pm$ 2.858 & 30.672 $\pm$ 2.551
  & 30.423 $\pm$ 2.871 & 37.546 $\pm$ 2.881 & 37.446 $\pm$ 1.595 \\
  & Deep Ens. & 30.090 $\pm$ 3.266 & 38.032 $\pm$ 1.804
  & 30.051 $\pm$ 2.834 & 35.965 $\pm$ 3.217 & 38.866 $\pm$ 1.732 \\
  & Im2Im-UQ    & 26.867 $\pm$ 2.923 & 33.135 $\pm$ 2.543 &
  32.739 $\pm$ 3.535 & 36.163 $\pm$ 3.059 & 25.833 $\pm$ 0.734 \\
  & Im2Im-Deep  & 29.711 $\pm$ 3.138 & 34.565 $\pm$ 2.393
  & 33.557 $\pm$ 4.018 & 37.062 $\pm$ 2.968 & 34.038 $\pm$ 1.837 \\
& Im2Im-Median & 29.744 $\pm$ 3.130 & 35.191 $\pm$ 2.398
& 33.708 $\pm$ 4.244 & 36.576 $\pm$ 2.824& 33.388 $\pm$ 3.646 \\
  & QUTCC       & 29.833 $\pm$ 3.156 & 34.948 $\pm$ 2.436
  & 33.660 $\pm$ 4.143 & 37.498 $\pm$ 3.797 & 37.350 $\pm$ 1.936 \\
\addlinespace

\multirow{6}{*}{LPIPS ($\downarrow$)}
  & MC-Drop.  & 0.345 $\pm$ 0.034 & 0.125 $\pm$ 0.018
  & 0.392 $\pm$ 0.122 & 0.191 $\pm$ 0.087 & 0.153 $\pm$ 0.014 \\
  & Deep Ens. & 0.313 $\pm$ 0.053 & 0.095 $\pm$ 0.011
  & 0.407 $\pm$ 0.129 & 0.192 $\pm$ 0.088 & 0.159 $\pm$ 0.015 \\
  & Im2Im-UQ    & 0.343 $\pm$ 0.033 & 0.153 $\pm$ 0.025 &
  0.420 $\pm$ 0.092 & 0.294 $\pm$ 0.071 & 0.360 $\pm$ 0.033 \\
  & Im2Im-Deep  & 0.324 $\pm$ 0.043 & 0.125 $\pm$ 0.015 &
  0.414 $\pm$ 0.103 & 0.299 $\pm$ 0.071 & 0.297 $\pm$ 0.029 \\
& Im2Im-Median & 0.322 $\pm$ 0.040 & 0.121 $\pm$ 0.015
& 0.405 $\pm$ 0.097 & 0.304 $\pm$ 0.071 & 0.324 $\pm$ 0.027 \\
  & QUTCC       & 0.323 $\pm$ 0.040 & 0.121 $\pm$ 0.015 &
  0.408 $\pm$ 0.102 & 0.284 $\pm$ 0.072 & 0.312 $\pm$ 0.026 \\
\bottomrule
\end{tabular*}
\label{tbl:Mean_model_peformance}
\end{table}

\subsubsection{Quantile-Conditioned Model vs Quantile Calibration Scheme}
To what extent do the improvements come from the quantile-conditioned model itself versus the proposed quantile-space calibration procedure? To address this question, we take the 5th and 95th quantile from our quantile model and calibrate it using the same procedure as Im2Im. We observe the following mean interval lengths for Gaussian, Poisson, Real Noise, MRI, and QPI - 0.0604, 0.0482, 0.0375, 0.1071, and 0.0631. Our quantile-conditioned, quantile-calibrated model achieves smaller interval widths on Gaussian, Poisson, and Real Noise. Performance on QPI is comparable between the two approaches, while MRI shows slight improvements with Im2Im's calibration. Because the results are mixed, we conclude that both the quantile conditioning of the model and the quantile-space calibration procedure contribute value and are complementary components of our approach. 

\subsubsection{Quantile Crossing Performance}

\captionsetup[subtable]{font=small,labelfont=bf}

\sisetup{
  scientific-notation=true,
  exponent-product=\cdot,
  output-exponent-marker=\text{e},
  round-mode=figures,
  round-precision=3,
  table-number-alignment = center,
  text-series-to-math = true,
  propagate-math-font = true
}

\begin{table}[t]
\caption{Quantile crossing analysis.}
\centering
\begin{subtable}{0.48\textwidth}
\centering
\papertablefont
\setlength{\tabcolsep}{3pt}
\begin{tabular*}{\textwidth}{@{\extracolsep{\fill}} l S S S}
\toprule
\textbf{Task} & \textbf{Crossed Pix.} & \textbf{Total Pix.} & \textbf{Ratio} \\
\midrule
MRI        & \num{22}      & \num{1638400000}  & \num{1.34277344e-8}       \\
QPI        & \num{5}       & \num{262144000}   & \num{1.90734863e-8}       \\
Gaussian   & \num{1104238} & \num{3355443200}  & \num{0.000329088569}      \\
Poisson    & \num{3485}    & \num{3355443200}  & \num{1.03861094e-6}       \\
Real Noise & \num{33830}   & \num{1048576000}  & \num{0.0000322628021}     \\
\bottomrule
\end{tabular*}
\caption{QUTCC crossings for quantiles $0.1,0.2,\ldots,0.9$.}
\label{tbl:quantile_crossing}
\end{subtable}%
\hfill
\begin{subtable}{0.48\textwidth}
\centering
\papertablefont
\setlength{\tabcolsep}{3pt}
\begin{tabular*}{\textwidth}{@{\extracolsep{\fill}} l S S S}
\toprule
\textbf{Task} & \textbf{Im2Im} & \textbf{QUTCC} & \textbf{Total Pix.} \\
\midrule
MRI        & \num{0}    & \num{0}    & \num{204800000} \\
QPI        & \num{0}    & \num{0}    & \num{32768000}  \\
Gaussian   & \num{3911} & \num{3296} & \num{419430400} \\
Poisson    & \num{7}    & \num{247}  & \num{419430400} \\
Real Noise & \num{0}    & \num{0}    & \num{131072000} \\
\bottomrule
\end{tabular*}
\caption{Crossings for upper and lower bounds of Im2Im versus QUTCC.}
\label{tbl:quantile_crossing_im2im_vs_qutcc}
\end{subtable}
\end{table}

In section~\ref{Calibration_step} we describe the conformal calibration step, which is dependent on the quantile function being monotonic. 
To ensure the validity of the predicted quantiles, specifically to avoid the issue of quantile crossing, we quantified the number of quantile crossing occurrences between $q = [0.1, 0.2, 0.3, ..., 0.9]$ in \ours (Tbl \ref{tbl:quantile_crossing}). 
Quantile crossing can undermine the interpretability of our uncertainty estimates, as it contradicts the notion that quantile functions should be non-decreasing/non-overlapping. The results indicate that across all imaging tasks, the ratio of quantile crossing occurrences is minimal. 

We additionally compare the quantile crossing occurrences between Im2Im-UQ and QUTCC, utilizing each model's respective bounds, which can be visualized in Tbl \ref{tbl:quantile_crossing_im2im_vs_qutcc}, observing comparable performance between both models.

\subsubsection{Post-Processing Monotonicity Constraint}
Because QUTCC calibrates prediction sets through a binary search over quantile levels, the procedure is justified when the learned quantile function is monotonic. When this assumption holds, the calibrated bounds returned by the binary search procedure are guaranteed to satisfy the $\alpha$ coverage level. However, if quantile crossings do occur during calibration, the validity of the resulting calibration can be compromised. We therefore include a monotonic post-processing variant that guarantees valid intervals. For each candidate pair $(q_{\mathrm{lo}}, q_{\mathrm{hi}})$ evaluated during calibration, we compute the raw predictions $Q_\theta(x,q_{\mathrm{lo}})$ and $Q_\theta(x,q_{\mathrm{hi}})$ and form the deployed interval using the pixelwise repair:
\[
L(x)=\min\{Q_\theta(x,q_{\mathrm{lo}}),Q_\theta(x,q_{\mathrm{hi}})\}, \qquad
U(x)=\max\{Q_\theta(x,q_{\mathrm{lo}}),Q_\theta(x,q_{\mathrm{hi}})\}.
\]
The empirical lower- and upper-tail risks used for calibration are then computed from these repaired endpoints, and the same repair is applied at test time. Because the repair is included inside the evaluation loop, the resulting bounds are guaranteed to satisfy the per-side coverage budget for the repaired intervals. We additionally monitor empirical risk monotonicity along the calibration path; if crossings or risk-direction violations indicate that binary search is unstable, calibration can fall back to a repaired grid search over candidate quantile pairs. In practice, as shown in Table~\ref{tab:monotonic_post_processing}, we find that quantile crossings occur sufficiently rarely that this refinement step has a negligible effect on the calibrated bounds, and the simpler calibration scheme without the monotonicity post-processing yields identical coverage and interval widths.
\begin{table}[t]
\centering
\caption{Comparison of calibrated quantiles with and without monotonic post-processing across all tasks.}
\papertablefont
\setlength{\tabcolsep}{3pt}
\begin{tabular*}{0.85\textwidth}{@{\extracolsep{\fill}}lcccc@{}}
\toprule
\multirow{2}{*}{\textbf{Task}} & \multicolumn{2}{c}{\textbf{Normal Calibration}} & \multicolumn{2}{c}{\textbf{Monotonic Post-Processing}} \\
\cmidrule(lr){2-3} \cmidrule(lr){4-5}
& \textbf{Lower $q$} & \textbf{Upper $q$} & \textbf{Lower $q$} & \textbf{Upper $q$} \\
\midrule
Gaussian   & 0.0078125 & 0.9794922 & 0.0078125 & 0.9794922 \\
Poisson    & $4.77{\times}10^{-7}$ & 0.9999995 & $4.77{\times}10^{-7}$ & 0.9999995 \\
Real-Noise & 0.2177734 & 0.9580078 & 0.2177734 & 0.9580078 \\
MRI        & 0.0380859 & 0.9570313 & 0.0380859 & 0.9570313 \\
QPI        & 0.0205078 & 0.9921875 & 0.0205078 & 0.9921875 \\
\bottomrule
\end{tabular*}
\label{tab:monotonic_post_processing}
\end{table}

\subsubsection{Quantile Embedding Ablation}
To validate our design choice of conditioning every residual block of our model with quantile embeddings, we evaluate the impact of restricting these embeddings to specific network components. We train six variants of our model on the Gaussian denoising task, where each variant keep the same architecture and optimization but restricts conditioning to a subset of the residual blocks. We test four competing hypotheses regarding where quantile information is more critical: (i) that the model only requires embedded feature extraction in the encoder; (ii) that the model only requires embeddings in the decoder; (iii) that the model only requires semantic conditioning at the bottleneck; and (iv) that the model only requires conditioning at the highest-resolution layers (input and/or output).

We report the mean interval length and risk coverage of each calibrated model, targeting $\alpha=0.1$, in Table~\ref{tbl:quantile_embedding_ablation}. The baseline configuration achieves the tightest intervals while maintaining valid risk coverage. Notably, restricting conditioning to the early layers or exclusively to the high-resolution layers results in poor calibration, suggesting that the model is unable to propagate quantile-specific information to the final output. Conversely, the model calibrates to a valid $\alpha$ but produces overly conservative intervals when conditioning on only the bottleneck or decoder. This suggests that the network learns more efficient features and can use them more effectively when quantile information is injected at multiple scales throughout the architecture.
% \begin{table}[h]
%     \caption{\textbf{Quantile Embedding Ablation}: We evaluate the impact of quantile conditioning at different stages of our U-Net architecture on the Gaussian denoising task. Our best-performing baseline model applies the embedding to every residual block. We report the mean interval length and calibrated risk targeting $\alpha=0.1$.}
%     \centering
%     %\rowcolors{1}{red!15}{red!15} % every row gets the same red background
%     \begin{tabular}{lcc}
%         \toprule
%         \textbf{Model} & \textbf{Interval Length} & \textbf{Total Risk} \\
%         \midrule
%         \textbf{Embedding in all blocks (ours)} & \textbf{0.0592} & \textbf{0.0908} \\
%         Embedding in first block only        & 0.0619 & 0.1243 \\
%         Embedding in encoder blocks only     & 0.0645 & 0.1199 \\
%         Embedding in middle block only       & 0.0695 & 0.0829 \\
%         Embedding in decoder blocks only     & 0.0679 & 0.0895 \\
%         Embedding in final block only        & 0.0648 & 0.0987 \\
%         Embedding in first and final blocks  & 0.0676 & 0.1012 \\
%         \bottomrule
%     \end{tabular}
%     \label{tbl:quantile_embedding_ablation}
% \end{table}

\begin{table}[t]
    \caption{Quantile embedding ablation on Gaussian denoising.}
    \centering
    \papertablefont
    \setlength{\tabcolsep}{3pt}
    \begin{tabular*}{0.75\textwidth}{@{\extracolsep{\fill}}lcc@{}}
        \toprule
        \textbf{Model} & \textbf{Interval Length} & \textbf{Total Risk} \\
        \midrule
        \textbf{Embedding in all blocks (ours)} & \textbf{0.0592} & \textbf{0.0908} \\
        Embedding in first block only        & 0.0619 & 0.1243 \\
        Embedding in encoder blocks only     & 0.0645 & 0.1199 \\
        Embedding in middle block only       & 0.0695 & 0.0829 \\
        Embedding in decoder blocks only     & 0.0679 & 0.0895 \\
        Embedding in final block only        & 0.0648 & 0.0987 \\
        Embedding in first and final blocks  & 0.0676 & 0.1012 \\
        \bottomrule
    \end{tabular*}

    \label{tbl:quantile_embedding_ablation}
\end{table}

\subsubsection{Mean Interval Length and Corresponding Risk}
\label{Supp:Mean_int_length}
We show QUTCC's performance gains stratified by pixel intensity in Table~\ref{tbl:model_comparison}. We demonstrate the mean performance in Table~\ref{table:interval_length_and_risk}. For all five tasks, QUTCC predicts smaller mean interval lengths while controlling the risk amongst the conformal methods. To provide additional reassurance that the calibration behaves as expected across different target risk levels, we evaluated one of our models, MRI, with $\alpha = 0.1$, $0.2$, $0.3$, and $0.4$. After conformal calibration, the empirical test risks were $0.10 \pm 0.03$, $0.20 \pm 0.04$, $0.30 \pm 0.04$, and $0.40 \pm 0.04$, respectively, closely matching the desired target risks.

\begin{table}[t]
\caption{Mean interval length and total risk after calibration.}
\centering
\papertablefont
\setlength{\tabcolsep}{2.5pt}
\begin{tabular*}{\textwidth}{@{\extracolsep{\fill}}llccccc@{}}
\toprule
{Metric} & {Method} & {MRI} & {QPI} & {Gaussian} & {Poisson} & {Real Noise}\\
\midrule
\multirow{4}{*}{Interval Length} 
& MC-Dropout & $0.185 \pm 0.022$ &  $0.062 \pm 0.003$& $0.084 \pm 0.022$ & $0.147 \pm 0.012$ & $0.030 \pm 0.004$\\

& Deep Ensemble & {0.1055 $\pm$ 0.059} & {0.041 $\pm$ 0.012} & {0.055 $\pm$ 0.046} & 0.044 $\pm$ 0.070 & {0.025 $\pm$ 0.026}\\

& Im2Im-Deep & 0.1096 $\pm$ 0.056 & 0.065 $\pm$ 0.007 & 0.063 $\pm$ 0.049 & 0.047 $\pm$ 0.038 & 0.038 $\pm$ 0.045\\
& Im2Im-Asym. & 0.1096 $\pm$ 0.048 & 0.064 $\pm$ 0.007 & 0.062 $\pm$ 0.028 & 0.046 $\pm$ 0.016 & 0.037 $\pm$ 0.007\\
& K-RCPS (k = 2) & 0.1084 $\pm$ 0.047 & 0.067 $\pm$ 0.006 & 0.060 $\pm$ 0.026 & 0.047 $\pm$ 0.014 & 0.038 $\pm$ 0.006\\
& QUTCC      & 0.1083 $\pm$ 0.057 & 0.063 $\pm$ 0.008 & 0.059 $\pm$ 0.048 & {0.040 $\pm$ 0.029} & 0.036 $\pm$ 0.035\\
\addlinespace
\multirow{4}{*}{Total-Risk} 
& MC-Dropout &  0.046 $\pm$ 0.081 &  $0.093 \pm 0.089$& $0.092 \pm 0.064$  & $0.087 \pm 0.161$ & $0.046 \pm 0.081$\\

& Deep Ensemble & 0.099 $\pm$ 0.034 & 0.097 $\pm$ 0.036 & 0.092 $\pm$ 0.034 & 0.073 $\pm$ 0.074 & 0.096 $\pm$ 0.018\\
& Im2Im-Deep & 0.098 $\pm$ 0.048 & 0.100 $\pm$ 0.100 & 0.094 $\pm$ 0.065 & 0.049 $\pm$ 0.042 & 0.096 $\pm$ 0.040\\
& Im2Im-Asym. & 0.099 $\pm$ 0.046 & 0.097 $\pm$ 0.097 & 0.099 $\pm$ 0.065 & 0.056 $\pm$ 0.045 & 0.098 $\pm$ 0.044\\
& K-RCPS (k = 2) & 0.103 $\pm$ 0.050 & 0.092 $\pm$ 0.094  & 0.103 $\pm$ 0.063 & 0.045 $\pm$ 0.041 & 0.074 $\pm$ 0.036\\
& QUTCC & 0.099 $\pm$ 0.035& 0.098 $\pm$ 0.099 & 0.090 $\pm$ 0.046 & 0.093 $\pm$ 0.091 & 0.098 $\pm$ 0.029\\
\bottomrule
\end{tabular*}
\label{table:interval_length_and_risk}
\end{table}

\begin{table}[t]
    \caption{Inference time for uncertainty-bound prediction. We report mean $\pm$ standard deviation in milliseconds.}
    \centering
    \papertablefont
    \setlength{\tabcolsep}{3pt}
    \begin{tabular*}{\textwidth}{@{\extracolsep{\fill}}lccccc@{}}
        \toprule
        \textbf{Task} & \textbf{MC-Drop.} & \textbf{Deep Ens.} & \textbf{Im2Im-UQ} & \textbf{Im2Im-Deep} & \textbf{QUTCC} \\
        \midrule
        Poisson & $4335.34 \pm 18.54$ & $840.11 \pm 0.83$ & $16.64 \pm 0.08$ & $84.00 \pm 0.14$ & $152.40 \pm 0.23$ \\
        Gaussian & $4346.23 \pm 1.07$ & $836.58 \pm 1.17$ & $16.52 \pm 0.09$ & $83.45 \pm 0.13$ & $151.72 \pm 0.24$ \\
        Real-Noise & $4343.69 \pm 3.15$ & $837.94 \pm 1.14$ & $16.58 \pm 0.21$ & $83.68 \pm 0.16$ & $152.16 \pm 0.35$ \\
        MRI & $1864.37 \pm 9.18$ & $359.14 \pm 4.15$ & $7.09 \pm 0.07$ & $35.30 \pm 0.13$ & $61.67 \pm 0.17$ \\
        QPI & $949.32 \pm 28.99$ & $185.23 \pm 10.77$ & $2.30 \pm 0.15$ & $14.69 \pm 0.65$ & $20.05 \pm 0.71$ \\
        \bottomrule
    \end{tabular*}
    \label{tab:inference_time}
\end{table}

\begin{table}[t]
    \caption{Training cost for each uncertainty method. We report the GPU hours for 50 training epochs.}
    \centering
    \papertablefont
    \setlength{\tabcolsep}{3pt}
    \begin{tabular*}{\textwidth}{@{\extracolsep{\fill}}lccccc@{}}
        \toprule
        \textbf{Task} & \textbf{MC-Drop.} & \textbf{Deep Ens.} & \textbf{Im2Im-UQ} & \textbf{Im2Im-Deep} & \textbf{QUTCC} \\
        \midrule
        Poisson & $52.11$ & $403.92$ & $4.28$ & $38.42$ & $39.96$ \\
        Gaussian & $52.92$ & $401.11$ & $4.32$ & $38.48$ & $39.50$ \\
        Real-Noise & $53.68$ & $533.66$ & $6.36$ & $55.63$ & $53.34$ \\
        MRI & $56.37$ & $411.80$ & $5.49$ & $41.17$ & $41.08$ \\
        QPI & $87.79$ & $590.31$ & $11.65$ & $56.71$ & $59.33$ \\
        \bottomrule
    \end{tabular*}
    \label{tab:training_time}
\end{table}

% Runtimes are standardized 50-epoch GPU-hour estimates from controlled timing runs on one NVIDIA RTX A6000 GPU. Values include the observed six-epoch wall-clock runtime, including setup and epoch 1, plus the median wall-clock time over epochs 2--6 multiplied by the remaining 44 epochs. Deep Ensemble multiplies one member timing by 10 members.

\subsubsection{Computational Cost}
Tables~\ref{tab:inference_time} and~\ref{tab:training_time} summarize the inference and training costs of each uncertainty method. Inference times are measured as the mean and standard deviation over 50 timed runs after 5 warmup runs on a single NVIDIA RTX A6000 GPU. Input sizes are $1\times512\times512$ for Poisson, Gaussian, and Real-Noise, $1\times320\times320$ for MRI, and $2\times128\times128$ for QPI. Because MC-Dropout uses 50 stochastic forward passes and Deep Ensemble uses 10 members, their inference and training costs are much higher than other methods. Im2Im-UQ is fastest because it uses the lightest network architecture, while Im2Im-Deep and QUTCC use the same deeper backbone. \ours is slightly slower at inference than Im2Im-Deep because estimating the lower, mean and upper quantiles are done in three separate forward passes for \ours, whereas Im2Im-Deep evaluates them as a single pass.  we report training costs as GPU-hour estimates for 50 epochs on the same A6000 hardware in Table~\ref{tab:training_time}. \ours has training cost comparable to Im2Im-Deep across tasks, while achieving performance similar to or exceeding Deep Ensemble, which is consistently the most expensive method because it trains 10 independent predictors.

Additionally, it takes approximately 2000 seconds ($\approx$30~minutes) to calibrate the quantile bounds of a PDF with 11 quantiles (q = 0.05, , 0.15, ..., 0.85, 0.95). The calibration time for constructing a PDF is largely independent of the number of specified quantiles. Although multiple quantile bounds at different risk levels must be calibrated to form the full distribution, each calibration can be performed as an independent job. As a result, these calibrations can be executed in parallel, and the overall calibration time is determined by the longest individual calibration rather than by the total number of quantiles. However, increasing the number of quantiles yields a more detailed representation of the underlying distribution (Fig.~\ref{sup:increasing_quantiles}). The PDF can then be reconstructed by performing one forward pass per quantile. Once calibrated for a given dataset, the PDF quantile bounds are fixed and can be consistently applied to all images within the dataset.

\subsubsection{Size-Stratified Risk}
\begin{figure}[htbp!]
  \centering
  \includegraphics[width=\linewidth]{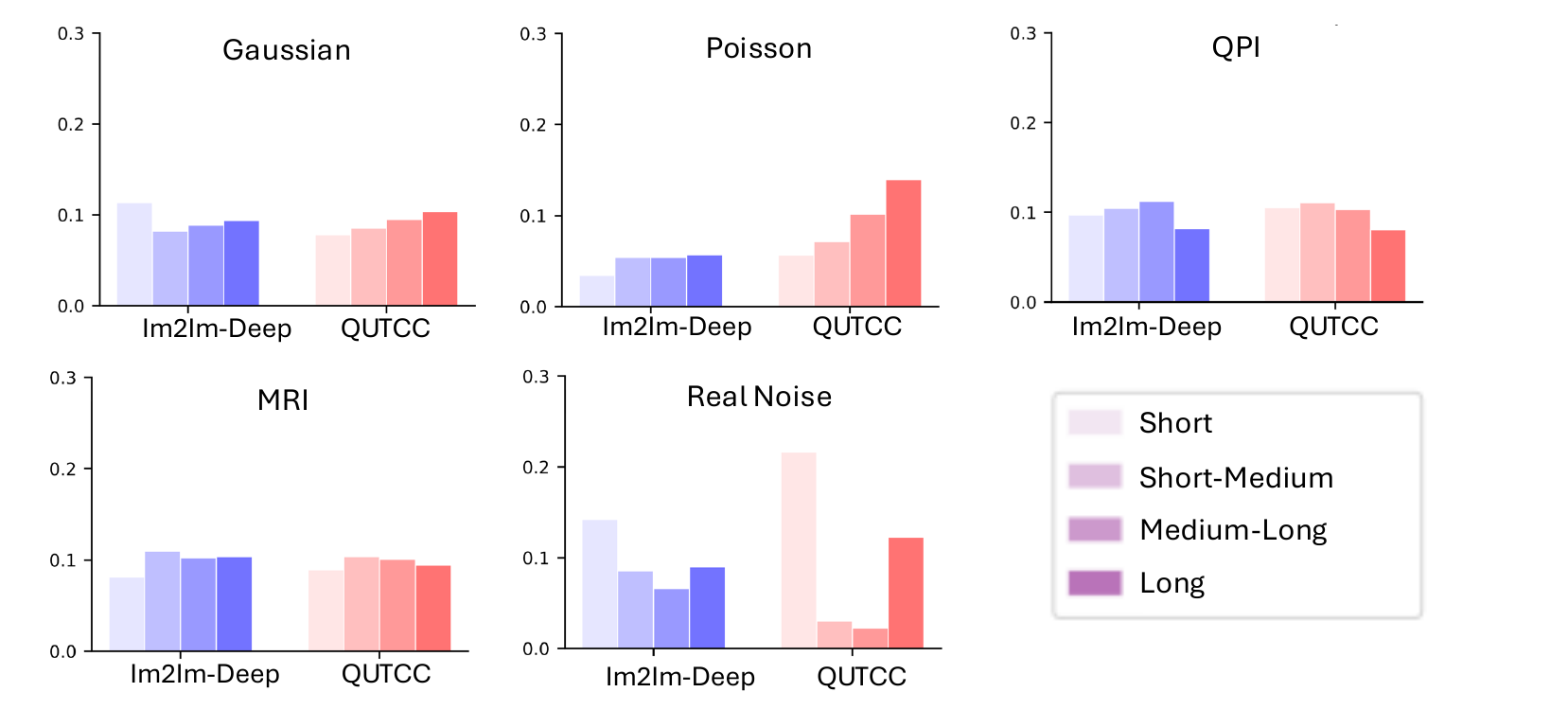}
  \caption{\textbf{Size-Stratified Risk of Im2Im-Deep vs. \ours}: We evaluate the size-stratified risk of Im2Im-Deep and \ours across all tasks. Overall, neither model exhibits a strong relationship between interval width and empirical risk, suggesting limited sensitivity to interval size. However, in the Gaussian and Poisson settings, both models display a mild trend toward improved calibration, or lower risk, for narrower prediction intervals. 
 }  
  \label{sup:size_stratified}
\end{figure}
%Size-stratified risk
We observed the size-stratified risk of all inverse tasks between Im2Im-Deep and \ours (Fig.~\ref{sup:size_stratified}). To calculate size-stratified risk, the prediction intervals are first binned into different sizes, ranging from smallest to largest. Then the risk is calculated across all the bins to ensure that the model’s uncertainty estimates are well-calibrated across different levels of confidence. While both Im2Im-Deep and \ours have bins that exceed the $\alpha$, generally, most bins fall under the chosen risk.

\subsubsection{Additional Visualizations and Hallucinations}
\label{sup:additional_visualizations}
\begin{figure}[htbp!]
  \centering
  \includegraphics[width=0.85\linewidth]{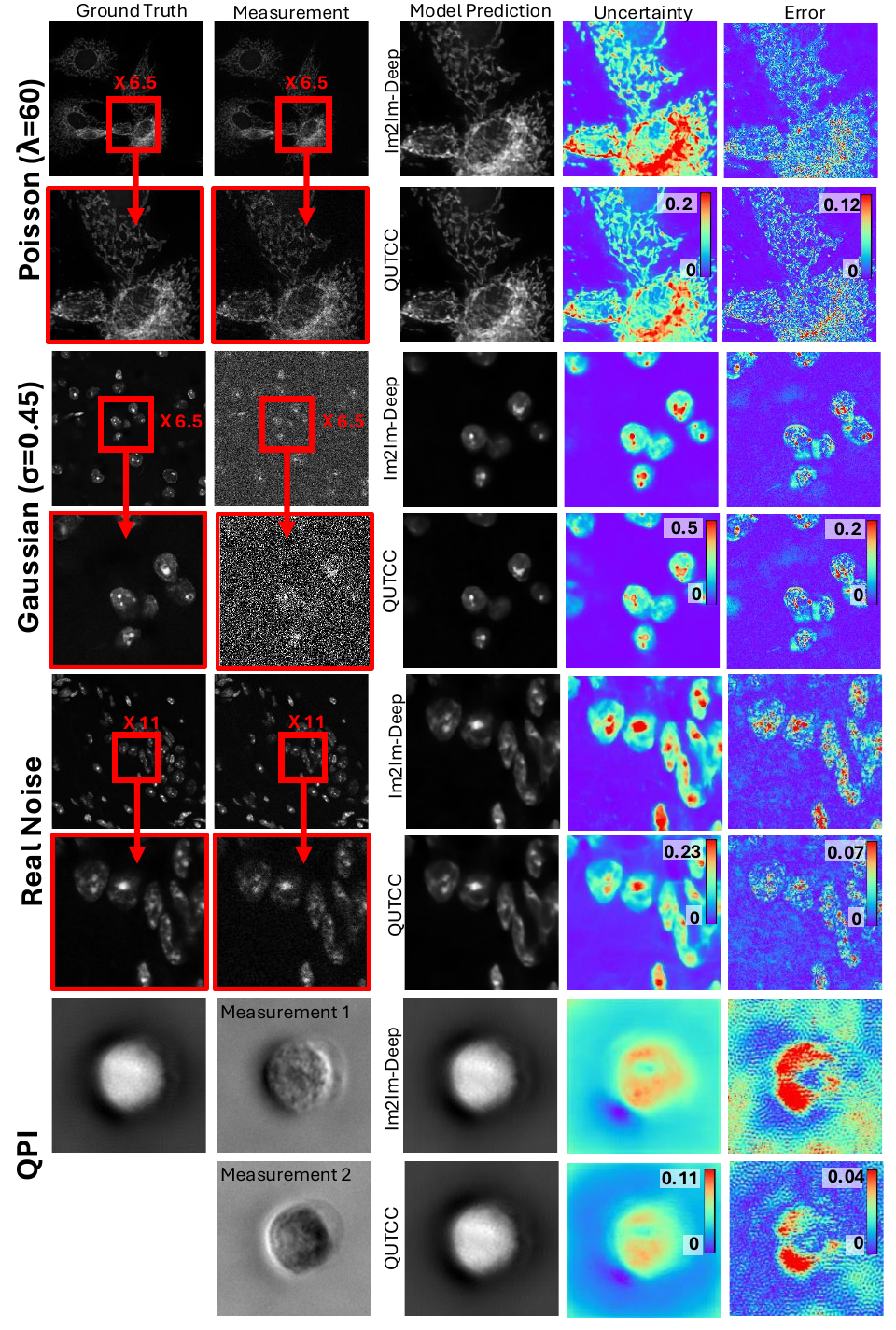}
  \caption{\textbf{Additional Uncertainty Visualizations:} We visualize both the full and zoomed-in regions of image reconstructions for QPI and denoising with Poisson, Gaussian and Real Noise. Consistent with observations presented in the results section, \ours produces more precise uncertainty estimates that closely align with localized regions of high reconstruction error. In contrast, Im2Im-Deep tends to highlight broader regions of uncertainty and lacks specificity, making it hard to distinguish areas of importance.  This highlights \ours~’s ability predict more informative uncertainty maps.
 }  
  \label{sup:supplement_visualizations}
\end{figure}

%Visualizations for all the modalities
We also provide visualizations of the remaining imaging tasks not included in the main results (Fig.~\ref{sup:supplement_visualizations}). For all sample tasks, both \ours and Im2Im-Deep effectively highlight regions with high reconstruction error. However, \ours has smaller interval lengths and produces slightly more localized uncertainty estimates, which may be more useful for downstream applications. 
%While Im2Im-Deep is capable of identifying regions of error, it tends to assign elevated uncertainty across larger portions of the sample, making it challenging for downstream tasks to prioritize regions based on uncertainty interval sizes. 
This trend is consistent across all five imaging inverse problems. 

\begin{figure}[htbp!]
  \centering
  \includegraphics[width=\linewidth]{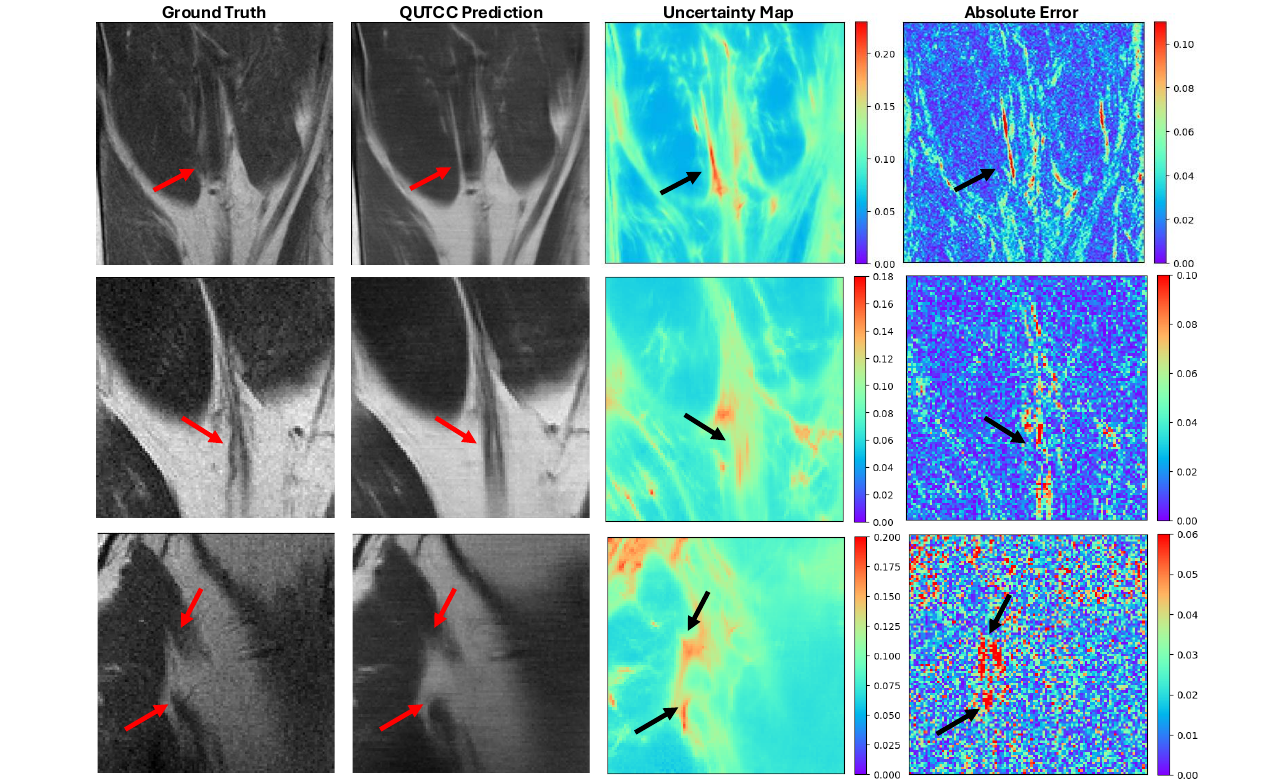}
    \caption{\textbf{\ours localizes MRI hallucinations.} We show additional MRI reconstruction results along with \ours's predicted uncertainty map and absolute reconstruction error. Red arrows mark structures that are hallucinated in the reconstruction relative to the ground truth. Black arrows show that \ours can produce precise uncertainty maps where high uncertainty is correlated with areas of high reconstruction error.}
  \label{sup:supplement_hallucinations}
\end{figure}

Figure~\ref{sup:supplement_hallucinations} provides additional MRI examples in which the reconstruction contains hallucinations that are not present in the ground truth. These errors are difficult to identify from the reconstructed image alone because the hallucinated features appear visually plausible. However, the \ours uncertainty predictions can help us locate suspicious regions with high uncertainty, which are often co-located with regions with high reconstruction error where hallucinations are more common.


\subsection{Bound Asymmetry and non-uniform scaling}
%Asymmetric bounds
\label{sec:adaptive_scaling_vs_krcps}
\begin{figure}[h!]
  \centering
  \includegraphics[width=\linewidth]{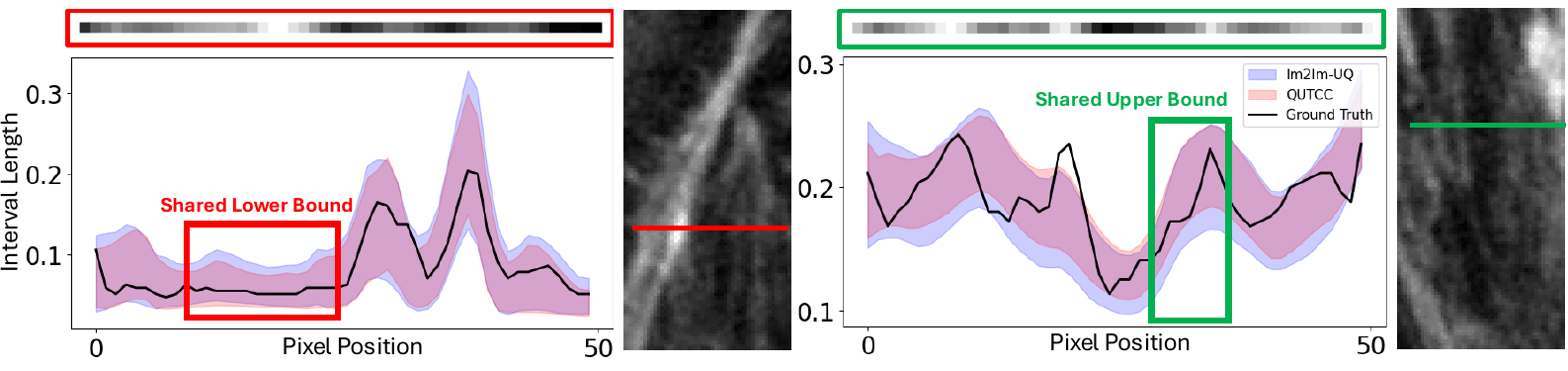}
  \caption{\textbf{\ours produces narrower intervals through asymmetric bounds}: We analyze the pixel-wise uncertainty bounds predicted by Im2Im-UQ and \ours and observe that \ours exhibits asymmetric behavior in its interval estimates. In Im2Im-UQ, both the upper and lower bounds are uniformly scaled by a global factor $\lambda$ to satisfy coverage constraints, which limits flexibility in adapting to signal-specific uncertainty. In contrast, \ours learns to predict quantiles directly, enabling it to independently modulate upper and lower bounds based on signal characteristics. This results in more adaptive and efficient uncertainty intervals. For instance, in the red boxed region, \ours matches Im2Im-UQ’s lower bound but produces a significantly tighter upper bound. Conversely, in the green boxed region, both models share an upper bound, yet \ours yields a tighter lower bound. Samples shown are Gaussian images with $\sigma = 0.1$.
 }  
  \label{sup:asymmetrical_bounds}
\end{figure}
How is \ours able to predict tighter uncertainty intervals while maintaining the same level of risk? In this section, we investigate the effects of asymmetry and non-uniform pixel-wise scaling. In Fig.~\ref{sup:asymmetrical_bounds} we investigate the effect of asymmetry. \ours produces asymmetric predictive intervals— its upper and lower bounds are adjusted independently. In contrast, Im2Im-UQ applies a single global scaling factor $\lambda$ uniformly to both bounds, which can be suboptimal in cases where only one side of the interval requires adjustment. In the red boxed region of Fig.~\ref{sup:asymmetrical_bounds}, both \ours and Im2Im-UQ share a similar lower bound, yet \ours predicts a significantly tighter upper bound. Similarly, in the green boxed region, both methods align on the upper bound, but \ours yields a tighter lower bound. These examples highlight \ours's ability to adaptively adjust its interval predictions, leading to more precise interval estimates.

%\subsection{Effect of Calibration Procedure on the Scaling Factor $\lambda$}

Next, we examine the effects of non-uniform, pixel-wise scaling. To do this, we demonstrate how scaling is applied during conformal calibration for Im2Im-UQ, Im2Im-UQ with K-RCPS calibration, and QUTCC. To ensure valid upper and lower bounds with statistical guarantees, the Im2Im calibration procedure measures miscoverage of the prediction intervals on a held-out calibration dataset. The bounds are then uniformly scaled by a constant factor $\lambda$, and the smallest $\lambda$ that controls the risk is selected. In the K-RCPS procedure, instead of calibrating a single global $\lambda$, pixels in the calibration dataset are grouped into membership classes (using Otsu’s method or any chosen heuristic), and a separate $\lambda$ is calibrated for each class. This $\lambda$ map is then applied on a dataset level. This behavior is illustrated in Fig.~\ref{sup:diff_lambdas}: the Im2Im $\lambda$ map is constant, whereas the K-RCPS $\lambda$ map contains two distinct values (for $k=2$ classes). 

In contrast, QUTCC does not explicitly calibrate $\lambda$. Instead, the calibration is a function of a neural network with input parameters, $q_{\text{hi}}$ and $q_{\text{lo}}$. This results in an implicit, pixel-wise scaling factor. While we do not predict a per-pixel $\lambda$ scaling, we can calculate the effective per-pixel scaling by comparing the initial and final calibrated bounds. The calibrated upper and lower bounds, $f_\theta(\mathbf y, \hat q_{\text{up}}), f_\theta(\mathbf y, \hat q_{\text{lo}})$, are related to the initial upper and lower bounds, $f_\theta(\mathbf y, \tilde q_{\text{up}}), f_\theta(\mathbf y, \tilde q_{\text{lo}})$, and the effective scaling, $\lambda_{\text{hi}}$, $\lambda_{\text{lo}}$, by the equations:
\begin{equation}
    f_\theta(\mathbf y, \hat q_{\text{lo}}) = \hat{\mathbf x} - \lambda_{\text{lo}}(f_\theta(\mathbf y, \tilde q_{\text{lo}}) - \hat{\mathbf x}), \quad f_\theta(\mathbf y, \hat q_{\text{hi}}) = \hat{\mathbf x} + \lambda_{\text{hi}}(f_\theta(\mathbf y, \tilde q_{\text{hi}}) - \hat{\mathbf x}).
\end{equation}

% \begin{equation}
%     l_c(\mathbf y) = \hat{\mathbf x} - \lambda_{\text{lo}}(l_i (\mathbf y) - \hat{\mathbf x}), \quad u_c(\mathbf y) = \hat{\mathbf x} + \lambda_{\text{hi}}(u_i (\mathbf y) - \hat{\mathbf x}).
% \end{equation}

Rearranging these equations to solve for the effective $\lambda$ as a function of the initial and final predicted bounds, we obtain:
\begin{equation}
    \lambda_{\text{hi}}(\mathbf y)[k] = \frac{f_\theta(\mathbf y, \hat q_{\text{hi}})[k] - f_\theta(\mathbf y, q=0.5)[k] }{f_\theta(\mathbf y, \tilde q_{\text{hi}})[k] - f_\theta(\mathbf y, q=0.5)[k] } , \quad \lambda_{\text{lo}}(\mathbf y)[k] = \frac{ f_\theta(\mathbf y, q=0.5)[k] - f_\theta(\mathbf y, \hat q_{\text{lo}})[k]}{f_\theta(\mathbf y, \tilde q_{\text{lo}}[k]) - f_\theta(\mathbf y, q=0.5) [k]},
\end{equation}
\noindent resulting in a per-pixel effective scaling that is adapted to each measurement. The corresponding upper and lower effective scaling maps, $\lambda_{\text{hi}}$ and $\lambda_{\text{lo}}$, for \ours are shown in Fig.~\ref{sup:diff_lambdas}. Here we can see that \ours automatically learns a pixel-wise scaling that depends on the features within the image, unlike Im2Im, which uses a constant, and K-RCPS, which uses a membership map based on a heuristic measure (e.g., intensity) on the dataset level. This per-pixel adaptivity enables QUTCC to produce more tailored uncertainty estimates, leading to tighter uncertainty intervals.
% \begin{figure}[]
%   \centering
%   \includegraphics[width=\linewidth]{Figures/Supplement_Figures/different_lambdas_supp.pdf}
%   \caption{\textbf{$\boldsymbol \lambda$ maps of Im2Im, K-RCPS, and QUTCC}: Im2Im-UQ learns a single global $\lambda$, K-RCPS calibrates class-wise $\lambda$ values, while QUTCC produces an implicit pixel-wise scaling through direct quantile calibration, enabling tighter and more adaptive uncertainty bounds.
%  }  
%   \label{sup:diff_lambdas}
% \end{figure}
\subsection{Additional PDF results}
\label{sup:additional_pdf}
%Bigger signal vs smaller signal
\begin{figure}
  \centering
  \includegraphics[width=\linewidth]{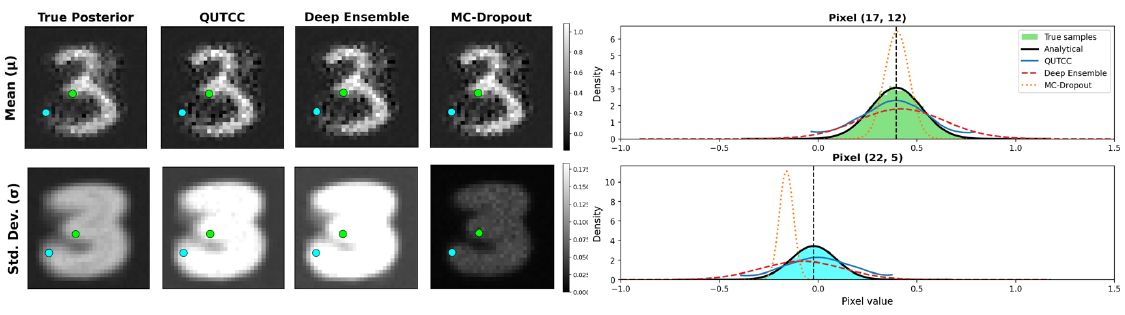}
    \caption{\textbf{\ours predicts more accurate pixel-wise PDFs of a known toy-distribution than compared to MC-Dropout and Deep Ensemble}} 
  \label{sup:toy_example}
\end{figure}

\begin{figure}
  \centering
  \includegraphics[width=.75\linewidth]{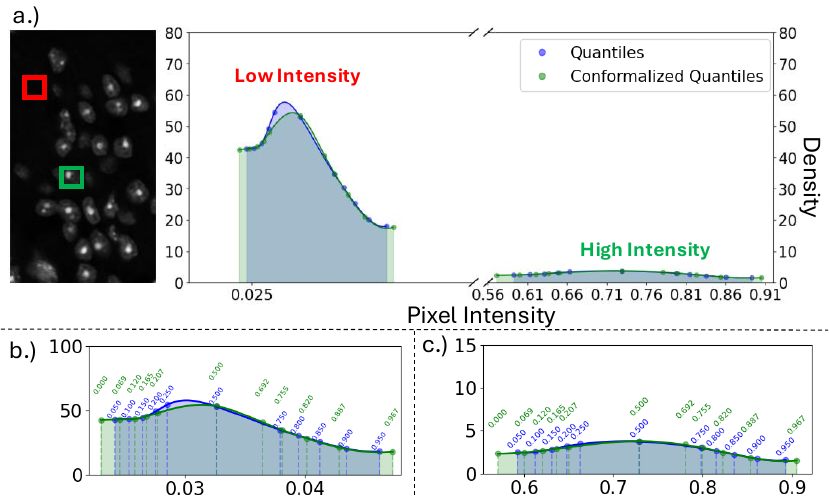}
    \caption{\textbf{\ours predicts different pixel-wise PDFs based on different signal intensity} \textbf{a)} Comparison of pixel-wise PDFs for representative low-intensity and high-intensity pixels in a Gaussian measurement ($\sigma = 0.4$). \textbf{b)} Detailed view of the low-intensity pixel PDF, exhibiting a narrow, high-density distribution concentrated around few intensity values, indicating low predictive uncertainty. \textbf{c)} Detailed view of the high-intensity pixel PDF, showing a broader, lower-density distribution with increased spread, reflecting higher predictive uncertainty in bright image regions. The blue lines show the uncalibrated model, while the green lines show the conformalized quantiles after calibration.} 
  \label{sup:high_low_pdf}
\end{figure}

In this section, we present additional results highlighting \ours's ability to estimate a conditional probability density function. In Fig. ~\ref{sup:toy_example}, we show the visual results of MC-dropout, Deep Ensemble, and \ours on a toy-distribution as described in the main text. In Fig.~\ref{sup:high_low_pdf}a, we show several of the predicted PDFs for image denoising with Gaussian noise with $\sigma = 0.4$. Detailed views of the corresponding pixel-wise PDFs are presented in Fig.~\ref{sup:high_low_pdf}b for the low-intensity pixel and Fig.~\ref{sup:high_low_pdf}c for the high-intensity pixel.
Each graph displays two PDFs: the blue PDF represents the quantile predictions from the uncalibrated model, while the green PDF represents the conformally calibrated quantiles that provide finite-sample statistical coverage guarantees. The conformal calibration procedure adjusts the quantile levels to ensure valid coverage properties. For instance, while the 25th and 75th percentiles theoretically provide $50\%$ coverage, conformal calibration determines that the 20.7th and 69.2nd percentiles are required to achieve statistically guaranteed $50\%$ coverage for this specific dataset and model. 
\begin{figure}
  \centering
  \includegraphics[width=0.7\linewidth]{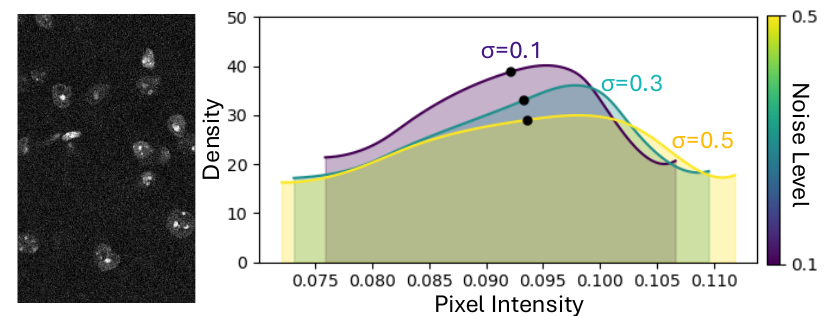}
  \caption{\textbf{PDF broadens as noise increases:} We observe the PDF of a single pixel under varying noise levels. At $\sigma = 0.1$, the noise is low, and the PDF is compact. As the noise increases to $\sigma = 0.3$ and $\sigma = 0.5$, the PDF gradually broadens, while the mean prediction value remains relatively unchanged. This broadening occurs due to the increased uncertainty introduced by higher noise levels. \ours successfully predicts this increased uncertainty as the noise increases. 
 }  
  \label{sup:diff_noise_pdf}
\end{figure}
\begin{figure}
  \centering
  \includegraphics[width=\linewidth]{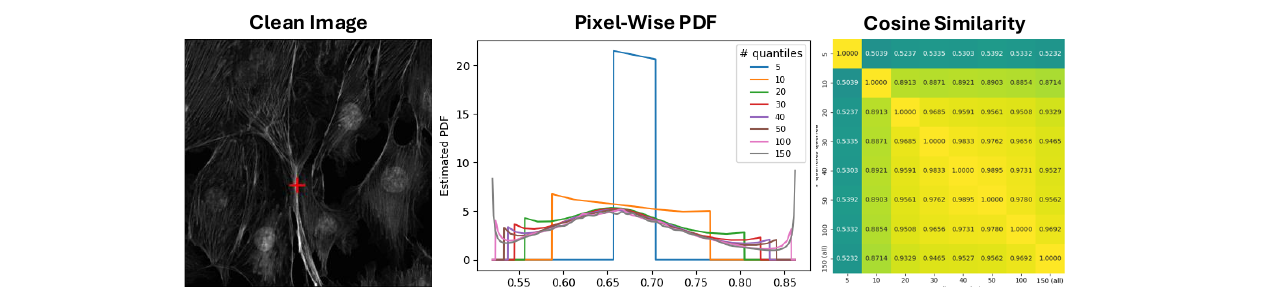}
  \caption{{\textbf{Pixelwise PDF as a function of queried quantiles}: We demonstrate the changes in PDF shape as the number of queried quantiles increase. We compare PDF similarity in a cosine similarity map.}}  
  \label{sup:increasing_quantiles}
\end{figure}
Additionally, Fig.~\ref{sup:diff_noise_pdf} illustrates the evolution of the pixel-wise PDF for a fixed pixel coordinate under varying Gaussian noise levels $\sigma \in \{0.1, 0.3, 0.5\}$. As the noise standard deviation increases, the PDF widens. This widening corresponds to increased epistemic uncertainty in the model's predictions, as higher noise levels reduce the information content available for accurate pixel intensity estimation. This then directly widens prediction intervals to maintain the desired coverage guarantees.
\subsubsection{Impact of Quantile number on estimated PDF}
We investigate how the number of quantiles influences the predicted PDF. The PDF is estimated from the numerical derivative of the quantile function. This estimate depends on the spacing between adjacent quantiles. Figure~\ref{sup:increasing_quantiles} illustrates this effect by showing PDFs reconstructed using progressively more quantile samples. The overall shape stays relatively similar as the number of quantile queries increases. This is quantitatively confirmed by measuring cosine similarity between the queried PDFs.
\subsection{Experiment Details}
\label{sup:experiment_details}
%Talk about each dataset that we used
%How many epochs we trained to
%What types of models
In this section, we provide additional experimental details about our training process and datasets used for our five different tasks: denoising (real, Gaussian, Poisson), MRI, and QPI. For Im2Im-Deep and \ours, we conducted a model selection sweep to identify the epoch that yielded the narrowest uncertainty intervals while satisfying the target risk level ($\alpha = 0.1$). The selected epochs are reported below for each task.
For all models, an initial learning rate of 1e-4 and weight decay of 0 was used. Batch size was adjusted depending on the task, with 4 used for denoising tasks, 12 for MRI, and 72 for QPI for U-Net backbones and 32, 512, and 16 (respectively) for ResNet-18 backbones. Images were normalized before training.
\subsubsection{Real Noise Task}
For the real noise task, we used the Fluorescence Microscopy Denoising (FMD) dataset \citep{zhang2019poisson}, which contains experimentally acquired fluorescence microscopy images spanning 12 wide-field, confocal, and two-photon modalities. The model was trained on 10,000 images, with 500 confocal mouse images used for calibration and an additional 500 for validation. The epochs used for Im2Im-Deep and \ours are 5 and 10, respectively. 
\subsubsection{Gaussian and Poisson Noise Task}
For both the Gaussian and Poisson tasks, we synthetically introduced varying levels of noise to the FMD ground truth images. The dataset was split into 180 ground truth images for training, 40 for calibration, and 20 for validation. The pseudocode for generating gaussian and poisson
noise are shown below in Algorithm \ref{alg:gaussian_noise} and Algorithm \ref{alg:poisson_noise}. 
\begin{figure}[h]
\centering
\begin{minipage}{0.48\textwidth}
\begin{algorithm}[H]
\caption{Add Gaussian Noise}
\label{alg:gaussian_noise}
\begin{algorithmic}[1]
\REQUIRE Clean image $\mathbf{x}$, max noise level $\sigma_{\text{max}}$
\ENSURE Noisy image $\mathbf{y}$
\STATE Sample noise std: $\sigma \sim \mathcal{U}(0, \sigma_{\text{max}})$
\STATE Sample Gaussian noise: $\boldsymbol{\eta} \sim \mathcal{N}(0, \sigma^2)$
\STATE Add noise: $\mathbf{y} \leftarrow \mathbf{x} + \boldsymbol{\eta}$
\STATE Clamp: $\mathbf{y} \leftarrow \text{Clamp}(\mathbf{y}, 0, 1)$
\STATE \textbf{return} $\mathbf{y}$
\end{algorithmic}
\end{algorithm}
\end{minipage}
\hfill
\begin{minipage}{0.48\textwidth}
\begin{algorithm}[H]
\caption{Add Poisson Noise}
\label{alg:poisson_noise}
\begin{algorithmic}[1]
\REQUIRE Clean image $\mathbf{x}$, min/max noise levels $\sigma_{\text{min}}, \sigma_{\text{max}}$
\ENSURE Noisy image $\mathbf{y}$
\STATE Sample scale: $\lambda \sim \mathcal{U}(\sigma_{\text{min}}, \sigma_{\text{max}})$
\STATE Scale image: $\mathbf{x}_{\text{scaled}} \leftarrow \lambda \cdot \mathbf{x}$
\STATE Sample noise: $\boldsymbol{\eta} \sim \text{Poisson}(\mathbf{x}_{\text{scaled}})$
\STATE Clamp: $\mathbf{y} \leftarrow \text{Clamp}(\boldsymbol{\eta}, 0, 1)$
\STATE \textbf{return} $\mathbf{y}$
\end{algorithmic}
\end{algorithm}
\end{minipage}
\end{figure}
For Gaussian noise, we set $\sigma_{\max} = 0.5$. For Poisson noise, the noise level range $(\lambda_{\min}, \lambda_{\max})$ was set to $(50, 100)$.  
In each iteration, with a batch size of 16, we generated 25 random noise levels uniformly sampled between the specified minimum and maximum values. The number of training epochs for Im2Im-Deep and \ours with Gaussian noise were 15 and 20, respectively. For Poisson noise, Im2Im-Deep was trained for 10 epochs, while \ours was trained for 35 epochs.

\subsubsection{Magnetic Resonance Imaging (MRI) Task}
Data used for the MRI task was obtained from the NYU fastMRI initiative database (fastmri.med.nyu.edu)~\citep{knoll2020fastmri,zbontar2018fastmri}. The primary goal of fastMRI is to test whether machine learning can aid in the reconstruction of medical MRI images. To train our models, we split the fastMRI dataset into 700 volumes for training, 200 volumes for calibration, and 200 for validation. It is important to note that a single volume contains multiple MRI images. The epochs used for Im2Im-Deep and \ours are 25 and 40, respectively.

To simulate the forward model in MRI, we start with a fully-sampled 3D volume composed of multiple 2D image slices. Each 2D image slice is transformed into its frequency domain representation using the 2D Fourier Transform, producing its k-space data. To simulate undersampled acquisition, we apply a 4$\times$ undersampling mask to the k-space. The resulting undersampled k-space is then transformed back into the image domain using the inverse Fourier Transform, yielding an aliased or artifact-corrupted image that serves as the input for the models.

\subsubsection{Quantitative Phase Imaging (QPI) Task}
Data used for the QPI task were obtained from the Berkeley Single Cell Computational Microscopy (BSCCM) dataset ~\citep{pinkard2024berkeley}. The BSCCM dataset contains image samples of individual white blood cells that have been captured with several illumination patterns on an LED array microscope. For this task, we used 289,059 images for training, 82,588 images for calibration, and 41,294 images for validation. The number of training epochs for Im2Im-Deep and \ours were 15 and 20, respectively.

In QPI, the goal is to image the structure of transparent cells by recovering the phase delay of the light that passes through the cells. Several intensity-only images are used to computationally estimate the phase of the light, since this cannot be measured directly. The input measurement $\mathbf{y}$ is the concatenation of two cell intensity images acquired at different illumination angles. The corresponding ground truth is the quantitative phase image recovered from four or more illumination angles. The model is trained to map these two intensity images into a phase image. 

 \subsection{LLM Usage}
LLMs were used to edit and correct grammar during writing, but were not involved in the research ideation process.

% \section{Technical appendices and supplementary material}
% Technical appendices with additional results, figures, graphs, and proofs may be submitted with the paper submission before the full submission deadline (see above). You can upload a ZIP file for videos or code, but do not upload a separate PDF file for the appendix. There is no page limit for the technical appendices. 

% Note: Think of the appendix as ``optional reading'' for reviewers. The paper must be able to stand alone without the appendix; for example, adding critical experiments that support the main claims to an appendix is inappropriate. 

%%%%%%%%%%%%%%%%%%%%%%%%%%%%%%%%%%%%%%%%%%%%%%%%%%%%%%%%%%%%

\newpage
\section*{NeurIPS Paper Checklist}

\begin{enumerate}

\item {\bf Claims}
    \item[] Question: Do the main claims made in the abstract and introduction accurately reflect the paper's contributions and scope?
    \item[] Answer:  \answerYes{}.
    \item[] Justification: In the abstract we claim  that our model produces tighter uncertainty intervals than prior methods, predicts plausible conditional distributions, and that high regions of uncertainty can be co-located with regions of hallucination. We support this claim in \textbf{Table 1}, \textbf{Figure 4}, and \textbf{Figure 5 }of the main paper.

\item {\bf Limitations}
    \item[] Question: Does the paper discuss the limitations of the work performed by the authors?
    \item[] Answer: \answerYes{}
    \item[] Justification: We provide limitations of our work in \textbf{Section 5: Limitations and Conclusions} of the main paper. We point out that our method is limited by a need for paired data and that the calibration procedure only provides pixel-wise marginal coverage guarantees. 
    
\item {\bf Theory assumptions and proofs}
    \item[] Question: For each theoretical result, does the paper provide the full set of assumptions and a complete (and correct) proof?
    \item[] Answer: \answerNA{} % Replace by \answerYes{}, \answerNo{}, or \answerNA{}.
    \item[] Justification:

    \item {\bf Experimental result reproducibility}
    \item[] Question: Does the paper fully disclose all the information needed to reproduce the main experimental results of the paper to the extent that it affects the main claims and/or conclusions of the paper (regardless of whether the code and data are provided or not)?
    \item[] Answer: \answerYes{}
    \item[] Justification: We fully disclose all the information needed to reproduce the main experimental results of the paper. 
    \begin{itemize}
        \item \textbf{Training/Testing Image Data:} We use only open-source data and these have been cited accordingly in the paper. We also elaborate more on experiment details in \textbf{Sec.~\ref{sup:experiment_details}}.
        
        \item \textbf{Model Architecture:} We describe our model architecture in the \textbf{Methodology Section}. We elaborate more in \textbf{Sec.~\ref{sup:model_arch}}.
    \end{itemize}

\item {\bf Open access to data and code}
    \item[] Question: Does the paper provide open access to the data and code, with sufficient instructions to faithfully reproduce the main experimental results, as described in supplemental material?
    \item[] Answer: \answerYes{}
    \item[] Justification: We provide an anonymous open source repository in the appendix: textbf{https://anonymous.4open.science/r/QUTCC-UQ-CB0F}. Additionally, we cite all datasets used in \textbf{Sec. A.6: Experiment Details}.

\item {\bf Experimental setting/details}
    \item[] Question: Does the paper specify all the training and test details (e.g., data splits, hyperparameters, how they were chosen, type of optimizer) necessary to understand the results?
    \item[] Answer:  \answerYes{}
    \item[] Justification: We fully disclose all the information needed to reproduce the main experimental results of the paper. We also elaborate more on experiment details in \textbf{Sec.~\ref{sup:experiment_details}}.

\item {\bf Experiment statistical significance}
    \item[] Question: Does the paper report error bars suitably and correctly defined or other appropriate information about the statistical significance of the experiments?
    \item[] Answer: \answerYes{}
    \item[] Justification: Yes, when reporting mean uncertainty length, MSE, SSIM, PSNR, and LPIPS, all values are reported with standard deviation values. 
    
\item {\bf Experiments compute resources}
    \item[] Question: For each experiment, does the paper provide sufficient information on the computer resources (type of compute workers, memory, time of execution) needed to reproduce the experiments?
    \item[] Answer: \answerYes{}
    \item[] Justification: We report the training cost for each uncertainty method on each image tasks in \textbf{Table 8} of the Appendix. We also disclose computational cost in \textbf{Sec. A.3.6}.

\item {\bf Code of ethics}
    \item[] Question: Does the research conducted in the paper conform, in every respect, with the NeurIPS Code of Ethics \url{https://neurips.cc/public/EthicsGuidelines}?
    \item[] Answer: \answerYes{}
    \item[] Justification: We do not have human subjects or participants. We use publicly available datasets and respect the terms of the datasets. We do not anticipate harmful consequences from this research.

\item {\bf Broader impacts}
    \item[] Question: Does the paper discuss both potential positive societal impacts and negative societal impacts of the work performed?
    \item[] Answer: \answerYes{} % Replace by \answerYes{}, \answerNo{}, or \answerNA{}.
    \item[] Justification: Both positive and negative societal impacts are discussed in the Conclusion.
    
\item {\bf Safeguards}
    \item[] Question: Does the paper describe safeguards that have been put in place for responsible release of data or models that have a high risk for misuse (e.g., pre-trained language models, image generators, or scraped datasets)?
    \item[] Answer: \answerNA{}
    \item[] Justification: \answerNA{}

\item {\bf Licenses for existing assets}
    \item[] Question: Are the creators or original owners of assets (e.g., code, data, models), used in the paper, properly credited and are the license and terms of use explicitly mentioned and properly respected?
    \item[] Answer: \answerNA{}
    \item[] Justification: \answerNA{}
    
\item {\bf New assets}
    \item[] Question: Are new assets introduced in the paper well documented and is the documentation provided alongside the assets?
    \item[] Answer: \answerNA{}
    \item[] Justification: \answerNA{}

\item {\bf Crowdsourcing and research with human subjects}
    \item[] Question: For crowdsourcing experiments and research with human subjects, does the paper include the full text of instructions given to participants and screenshots, if applicable, as well as details about compensation (if any)? 
    \item[] Answer: \answerNA{}
    \item[] Justification: \answerNA{}
    
\item {\bf Institutional review board (IRB) approvals or equivalent for research with human subjects}
    \item[] Question: Does the paper describe potential risks incurred by study participants, whether such risks were disclosed to the subjects, and whether Institutional Review Board (IRB) approvals (or an equivalent approval/review based on the requirements of your country or institution) were obtained?
    \item[] Answer: \answerNA{}
    \item[] Justification: \answerNA{}
    \item[] Guidelines:

\item {\bf Declaration of LLM usage}
    \item[] Question: Does the paper describe the usage of LLMs if it is an important, original, or non-standard component of the core methods in this research? Note that if the LLM is used only for writing, editing, or formatting purposes and does \emph{not} impact the core methodology, scientific rigor, or originality of the research, declaration is not required.
    %this research? 
    \item[] Answer: \answerNA{}
    \item[] Justification: \answerNA{}

\end{enumerate}

\end{document}